\documentclass[11pt,a4paper]{article}

 \pdfoutput=1
  
\usepackage{amsmath,amssymb,multirow}
\usepackage[affil-it]{authblk}
\usepackage[export]{adjustbox}

\usepackage{epsfig}  
\usepackage{graphicx}               % Standard graphics package  
\usepackage{url}
\usepackage[colorlinks=true,linkcolor=blue,citecolor=blue,urlcolor=blue]{hyperref}
\usepackage{cite}
\usepackage{caption}
\usepackage{subcaption}

\usepackage{color}

\usepackage{microtype}

\usepackage[margin=2.5cm]{geometry}

\title{  
\vspace*{-2.3cm}  
\begin{flushright}  
{\normalsize{  
CERN-PH-TH-2015-031, DESY 15-026
  }  }
\end{flushright}  
\vspace{2.5cm}  
\Huge  
\textbf{
Emerging Jets
}\vspace*{0.5cm}   
}

\author[a]{\bf Pedro Schwaller\thanks{pedro.schwaller@cern.ch}}
\affil[a]{CERN TH-PH Division, Meyrin, Switzerland}
\author[a]{\bf Daniel Stolarski\thanks{daniel.stolarski@cern.ch}}
\author[a,b]{{\bf Andreas Weiler}\thanks{andreas.weiler@cern.ch}}
\affil[b]{DESY, Notkestrasse 85, D-22607 Hamburg, Germany}

\date{}

\begin{document}  

\setcounter{page}{0}  
\maketitle  

\vspace*{0.5cm}  
\begin{abstract} 
In this work, we propose a novel search strategy for new physics at the LHC that utilizes calorimeter jets that (i) are composed dominantly of displaced tracks and (ii) have many different vertices within the jet cone. Such \textit{emerging jet} signatures are smoking guns for models with a composite dark sector where a parton shower in the dark sector is followed by displaced decays of dark pions back to SM jets. No current LHC searches are sensitive to this type of phenomenology. We perform a detailed simulation for a benchmark signal with two regular and two emerging jets, and present and implement strategies to suppress QCD backgrounds by up to six orders of magnitude. At the 14~TeV LHC, this signature can be probed with mediator masses as large as 1.5~TeV for a range of dark pion lifetimes, and the reach is increased further at the high-luminosity LHC.  The emerging jet search is also sensitive to a broad class of long-lived phenomena, and we show this for a supersymmetric model with $R$-parity violation. Possibilities for discovery at LHCb are also discussed. 
\end{abstract}

\thispagestyle{empty}  
\newpage  
  
\setcounter{page}{1}  

\vspace{-3cm}

\setcounter{tocdepth}{1}
\tableofcontents   
 
\vspace{2cm}
  
\baselineskip18pt

%--------------------------------------------------------------------%
\section{Introduction}
\label{sec:intro}
%--------------------------------------------------------------------%

The LHC has begun its exploration of the TeV scale, but as yet it has not uncovered any evidence for physics beyond the Standard Model (SM). Because of the complicated nature of the data taken at the LHC, it is of crucial importance to understand all the possible new physics scenarios that could be discovered. Digging out physics beyond the SM is difficult, and if the experimenters do not know what they are looking for, it is possible that there is evidence for new physics in the current data which can be discovered if a targeted search is performed. In this paper, we will give an example of a new type of reconstruction object which current searches are insensitive to and motivate why the experimental collaborations should begin a search for these objects.

These new objects arise naturally in many models of dark matter. Dark matter is known to require physics beyond the SM, but searches for weakly interacting massive particles (WIMPs)~\cite{Feng:2010gw} have so far come up empty. Furthermore, there are several astrophysical anomalies which may point away from the standard cold dark matter picture and instead towards dark matter with large self interactions~\cite{Rocha:2012jg,Peter:2012jh,Vogelsberger:2012ku,Zavala:2012us}, possibly hinting at more complicated dark matter sectors. For example, if the dark matter arises from a confining hidden sector~\cite{Kribs:2009fy,Alves:2009nf,Falkowski:2009yz,Alves:2010dd,Feng:2011ik,Kumar:2011iy,Bai:2013xga,Cline:2013zca,Boddy:2014yra,Hur:2011sv,Detmold:2014qqa,Krnjaic:2014xza}, then it will naturally be self interacting.  Another puzzle of dark matter is the coincidence between the energy density of dark matter and baryons. This comes out accidentally in the WIMP paradigm but can be explained if the dark matter abundance arises as an asymmetry much like the baryon abundance in QCD. In particular, if the same physics generates both asymmetries~\cite{Nussinov:1985xr,Kaplan:1991ah,Barr:1990ca,Barr:1991qn,Dodelson:1991iv,Fujii:2002aj,Kitano:2004sv,Farrar:2005zd,Gudnason:2006ug,Kitano:2008tk,Kaplan:2009ag,Shelton:2010ta,Davoudiasl:2010am,Buckley:2010ui,Blennow:2010qp,Cohen:2010kn,Frandsen:2011kt} (for a review see~\cite{Petraki:2013wwa,Zurek:2013wia}), then there will be a portal from the SM to the dark sector, and the GeV scale will play an important role on both sides.

Many of the models with a shared asymmetry between dark matter and baryons explain the similarity between the number densities of the two species, but the GeV scale is put into the dark sector by hand giving rise to a new coincidence. Combining the ideas of a confining hidden sector and cogeneration of dark matter with baryons can lead to a scenario that explains the coincidence of both the mass and number density of dark matter and baryons~\cite{Bai:2013xga}.\footnote{For a model that uses a confining hidden sector to explain the galactic center gamma ray excess see~\cite{Freytsis:2014sua}.} In the models presented in~\cite{Bai:2013xga}, there is a dark gauge group, and new matter is introduced to relate the running of the QCD and dark gauge couplings such that their confinement scales are near one another at the GeV scale. The new matter is also needed to generate the asymmetry. Therefore, this new matter must be charged under QCD, and it ends up acting as a portal between the visible and dark sectors that is accessible to colliders if it is sufficiently light. The analysis of~\cite{Bai:2013xga} points to new matter at the TeV scale, making the LHC the ideal machine to explore this class of hidden sector models.

The lightest baryon in the hidden sector is stable in analogy with the proton, so it is a good dark matter candidate. The phenomenology of this sector, however, is much more interesting than the usual WIMP scenarios because of the zoo of particles that are unstable. In particular, the TeV scale fields cause the mesons of the dark sector to decay back to the SM. Because of the GeV to TeV hierarchy, the decay back into the SM can be quite slow, with dark mesons traveling macroscopic distances before decaying. This is the basis of the novel collider phenomenology we will explore. 

\begin{figure}
\centering
\includegraphics[width=0.7\textwidth]{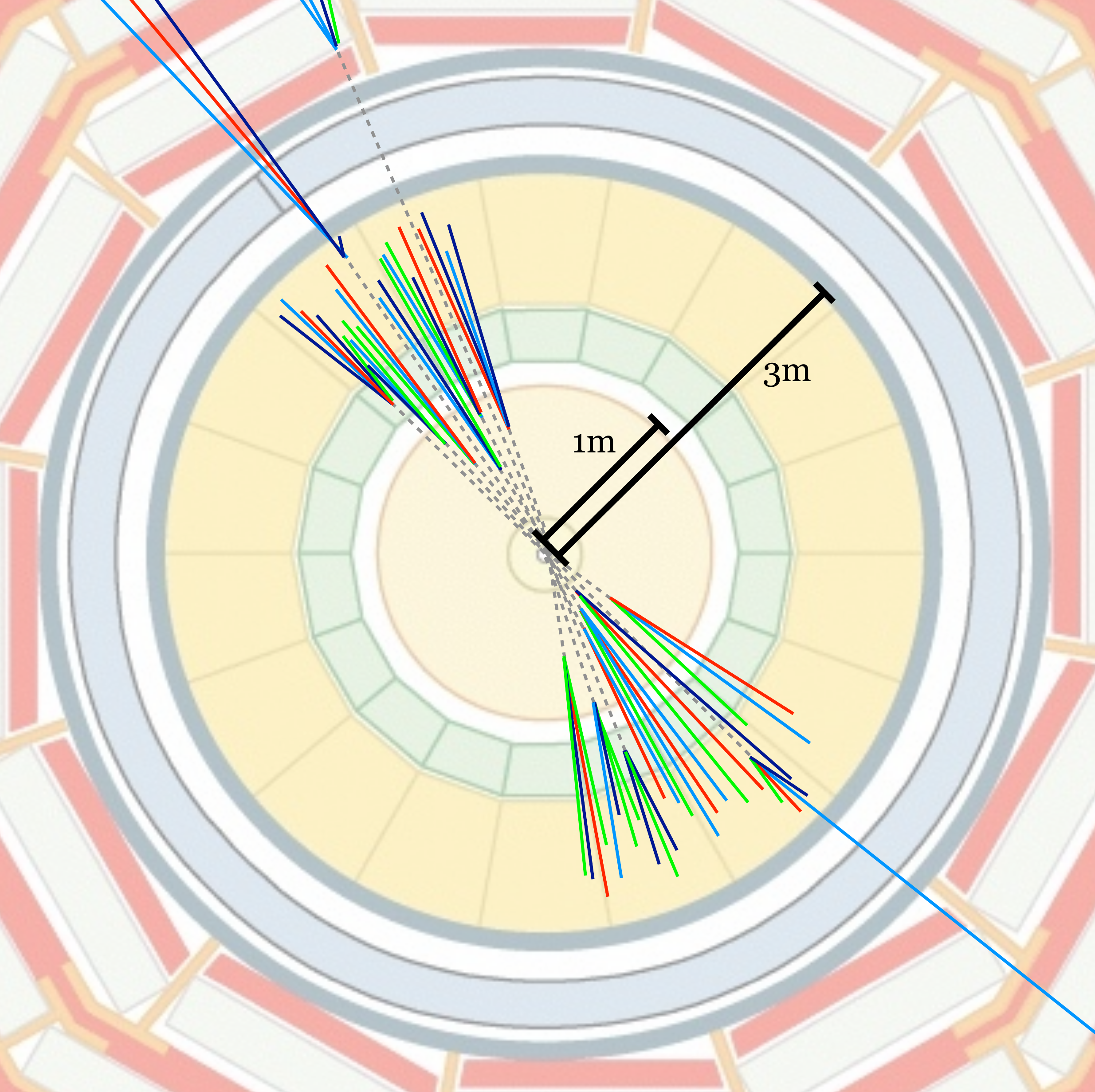}
\caption{A schematic depiction of pair production of dark quarks forming two emerging jets. Shown is an $x-y$ cross section of a detector with the beam pipe going into the page. The approximate radii of the tracker and calorimeter are also shown.  The dark mesons are represented by dashed lines because they do not interact with the detector. After traveling some distance, each individual dark pion decays into Standard Model particles, creating a small jet represented by solid colored lines. Because of the exponential decay, each set of SM particles originates a different distance from the interaction point, so the jet slowly emerges into the detector.  }
\label{fig:emerging-jets}
\end{figure}

Events from this type of scenario are shown schematically in Fig.~\ref{fig:emerging-jets} and can be described as follows. Consider the production of a TeV scale field which decays to two dark quarks and possibly other SM fields. The energy of these dark quarks will each be much larger than the confinement scale of the dark gauge group, so the dark quarks will shower and then hadronize producing a large number of dark mesons. If the dark sector is QCD like, then the dark hadrons will form into two jet-like structures, with all the hadrons going roughly in the same direction as one of the initial quarks. Motivated by the models in~\cite{Bai:2013xga}, we take the dark mesons to decay into SM quarks with a lifetime of order centimeters. Therefore, the dark jets will gradually turn into visible over a length scale of a few centimeters. Because of the exponential decay law, however, each hadron will decay in a different place in the detector and the jets will \textit{emerge} into the visible sector. 

Signals with jets of dark/hidden sector particles were considered before in the context of Hidden Valley models~\cite{Strassler:2006im}, and the possibility that at least some of the dark hadrons could decay with displaced vertices is discussed. Concrete proposals to search for hidden sectors~\cite{Strassler:2006im,Strassler:2006ri,Strassler:2006qa,Han:2007ae,Verducci:2011zz,Chan:2011aa} have however focussed on different aspects of hidden valley phenomenology. The possibility to use displaced vertices from individual dark pion decays  as background discrimination is mentioned in~\cite{Strassler:2006im,Strassler:2006ri}, but is not applicable to the signal proposed here with many overlapping displaced decays. Finally in~\cite{Strassler:2008fv} a scenario is discussed where dark hadrons decay promptly into heavy flavor quarks. This would lead to a large multiplicity of non-prompt tracks from the heavy flavor decays, but unlike in our scenario, the lifetimes of the hadrons are known and standard $b$ tagging technology can be used. Furthermore, light flavored mesons will also be produced in the hadronization of the heavy flavor pairs, so, unlike the signature proposed here, those events would also have many prompt tracks. 

%There are many other proposals to search for hidden sectors~\cite{Strassler:2006im,Strassler:2006ri,Strassler:2006qa,Han:2007ae,Verducci:2011zz,Chan:2011aa}, but as yet none which have explored this distinct signature. 
The main requirements for a model to produce emerging jet phenomenology are:
\begin{itemize}
\item A large hierarchy between the mediator mass and the hidden sector mass.
\item Strong coupling in the hidden sector so that there can be large particle multiplicity.
\item Macroscopic decay lengths of hidden sector fields back to the visible sector.
\end{itemize}
The purpose of this work is to characterize the emerging jets signature at hadron colliders and to develop an analysis strategy for the LHC experiments. In Sec.~\ref{sec:models}, we introduce the models which give rise to emerging jets and motivate the parameter space we consider, followed in Sec.~\ref{sec:ph} by a detailed description and modeling of emerging jet phenomenology and a discussion of existing searches. The emerging jet analysis strategy is detailed in Sec.~\ref{sec:analysis}, including simulations of signals and backgrounds. The projected reach at the 14~TeV LHC is shown in Fig.~\ref{fig:reachAB}. While the main analysis is based on reconstructing calorimeter jets with no prompt tracks, we also propose an alternative strategy using $p_T$ weighted tracks. We also outline a strategy for searching for emerging jet like signatures with the LHCb detector in Sec.~\ref{sec:lhcb}. 

While the analysis method presented here was designed with specific models in mind, these techniques are sensitive to a broad class of models with displaced phenomena. As an example, we show the reach for certain $R$-parity violating (RPV) supersymmetric scenarios in Sec.~\ref{sec:other}, finding excellent reach for this class of models shown in Fig.~\ref{fig:RPVreach}. Finally, in two appendices we present more details on the simulation of the signals and backgrounds, on the tracking algorithms, and a discussion of variations of model parameters and their impact on the analysis.

%--------------------------------------------------------------------%
\section{Models}
\label{sec:models}
%--------------------------------------------------------------------%

\begin{figure}
\centering
\begin{minipage}[c]{0.6\textwidth}
\includegraphics[width=0.9\textwidth]{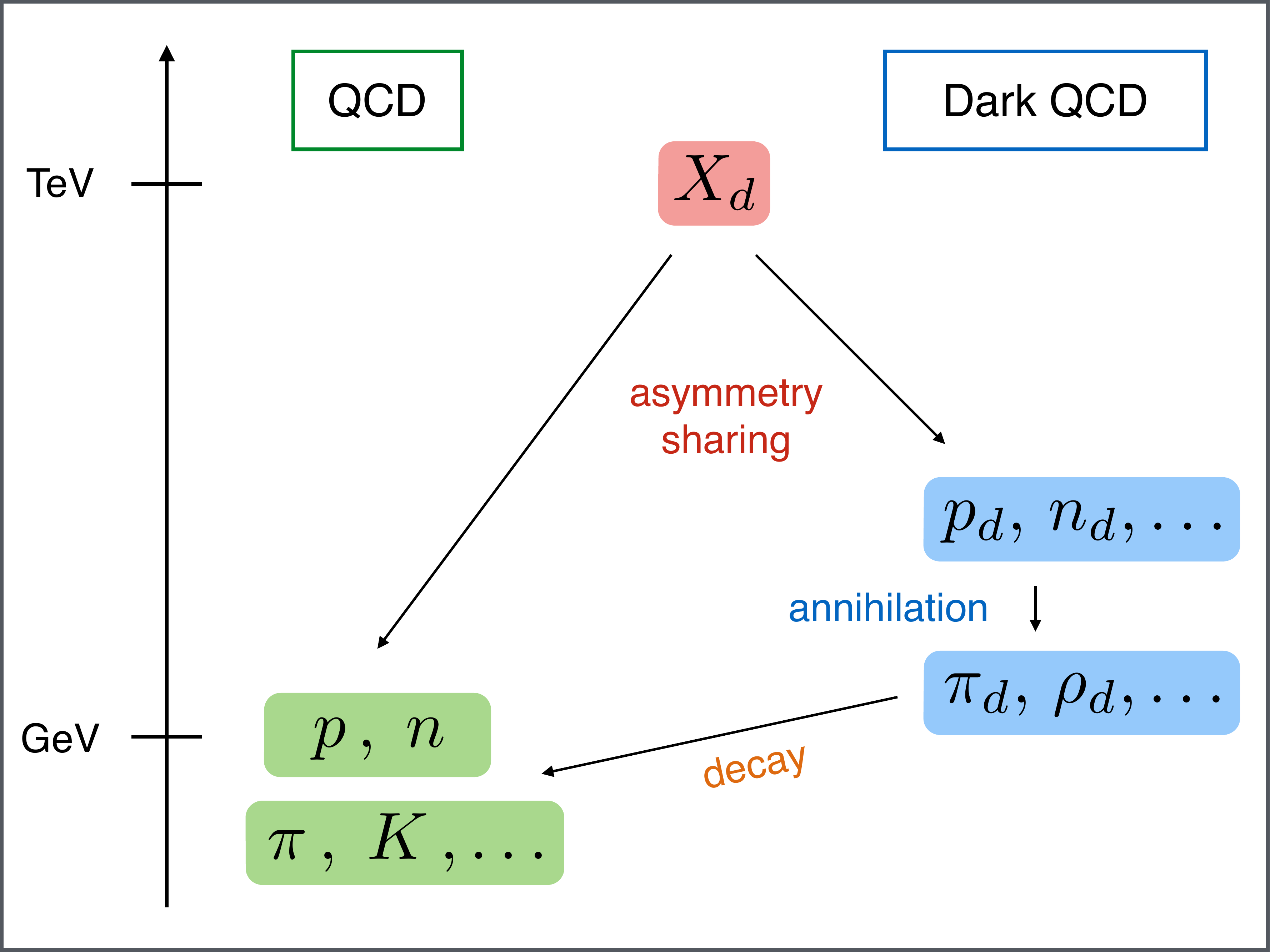}
\end{minipage}
\begin{minipage}[c]{0.39\textwidth}
\caption{Graphical representation of the dark QCD model. Baryon and dark matter asymmetries are shared via a mediator $X_d$ resulting in an asymmetry in the stable dark baryons $p_d$, $n_d$. The symmetric relic density is annihilated efficiently into dark pions, which eventually decay into SM particles. The DM number density is naturally of the same order as that of baryons, so the correct DM relic density is obtained when the dark baryon masses are in the 10~GeV range.\\}
\label{fig:model}
\end{minipage}
\end{figure}

We now describe our general setup which is shown schematically in Fig.~\ref{fig:model}. We consider a class of models with a dark sector with a non-abelian gauge symmetry, \textit{dark QCD}, that confines in the infrared, in a way similar to QCD. More concretely, we consider an extension of the standard model gauge group to %
\begin{align}
	 G_{\rm SM} \times SU(N_d)\,,
\end{align}
where $G_{\rm SM} =SU(3)_c \times SU(2) \times U(1)$ is the standard model gauge symmetry, and $N_d\geq 2$ is the number of dark colors. Furthermore we assume that there are $n_f$ Dirac fermions that are fundamentals of $SU(N_d)$ and singlets under $G_{\rm SM}$, which we refer to as \textit{dark quarks} $Q_d$. The dark sector confines at a scale $\Lambda_d$, which is the approximate mass of the majority of the dark mesons and baryons. The theory also contains pseudo-Goldstone bosons, analogous to QCD pions, which we take to have a common mass $m_{\pi_d}$ with  $m_{\pi_d} < \Lambda_d$. Motivated by asymmetric dark matter, we take the dimensionful parameters of the dark sector $\Lambda_d$ and $m_{\pi_d}$ to be $\mathcal{O}(1\, - \, 10)$ GeV.

The dark baryons carry a conserved charge, \textit{dark baryon number}, such that the lightest one is stable and constitutes the dark matter candidate of our model. On the other hand, the dark mesons do not carry such a conserved charge and can therefore decay to SM particles.

The dark sector is connected to the visible sector by a heavy mediator, making this similar in spirit to hidden valley models~\cite{Strassler:2006im}. Inspired by~\cite{Bai:2013xga}, we focus on a scalar mediator which is a bifundamental under both QCD and dark color. The bifundamental, $X_d$, can be pair produced and each one will decay to an SM quark and a dark quark.  Another possibility for a mediator is a neutral vector  $Z_d$ which couples to both quark pairs and dark quark pairs. The $Z_d$ is a nice toy model for studying dark sector properties, but we leave detailed studies of its phenomenology at the LHC to future work. The full particle content is summarized in Tab.~\ref{tab:fields}.

For the scalar mediator with the hypercharge assignment in Tab.~\ref{tab:fields}, the only allowed Yukawa type coupling is of the form~\cite{Bai:2013xga}
\begin{align}\label{eqn:darkYukawa}
	{\cal L}_\kappa = \kappa_{i j}\bar{Q}_{d_i} q_j X_d + {\rm h.c.}
\end{align}
where $q_j$ are the right-handed down-type SM quarks and $\kappa$ is a $n_f \times 3$ matrix of Yukawa couplings. Such couplings could in general lead to large flavor violating processes, but can be brought into agreement with experimental bounds if dark flavor originates from the same dynamics as the SM flavor structure or certainly if flavor symmetries are imposed on the dark sector~\cite{Brod:2014loa,Agrawal:2014aoa,Calibbi:2015sfa}. 
For definiteness, the fundamental Lagrangian which defines the model at high scales is given by
\begin{align}
	{\cal L} &\supset \bar{Q}_{d_i} (D\!\!\!\!\slash -m_{d_i}) Q_{d_i} + (D_\mu X_d ) (D^\mu X_d)^\dagger - M^2_{X_d} X_d X_d^\dagger -\frac{1}{4} G^{\mu \nu}_d G_{\mu\nu, d} +  {\cal L}_\kappa + {\cal L}_{\rm SM} \,,
\end{align}
where $G_d^{\mu \nu}$ is the dark gluon field strength tensor, and the covariant derivatives contain the couplings to the gauge fields. 

\begin{table}[t]
\begin{center}
\begin{tabular}{|c|c|c|c|c| }
\hline
 Field & $SU(3)\times SU(2) \times U(1)$ & $SU(3)_{\rm dark}$ & Mass & Spin \\\hline\hline
 $Q_d$ & $(1,1,0)$ & (3) & $m_d~{\cal O}(\rm GeV)$ & Dirac Fermion \\
 $X_d$ & $(3,1, \frac{1}{3})$ & (3) & $M_{X_d} ~{\cal O}(\rm TeV)$ & Complex Scalar \\ \hline
 $Z_d$ & $(1,1,0)$ & (1) & $M_{Z_d} ~{\cal O}(\rm TeV)$ & Vector Boson \\\hline
\end{tabular}
\end{center}
\caption{Particle content relevant for phenomenology. 
We use the $Z_d$ as a toy model and leave detailed study to future work.  }
\label{tab:fields}
\end{table}%

For the vector mediator, we assume that it couples vectorially to SM and dark quarks with couplings $g_{q}$ and $g_{d}$. While here we assume that $Z_d$ originates from a $U(1)$ symmetry broken at the TeV scale, it could in principle also originate from a non-abelian horizontal symmetry as in Ref.~\cite{Blennow:2010qp}, where the Sphaleron associated with this gauge interaction is used to connect the dark matter with the baryon asymmetry. 

\subsection{Mass Scales}

The present work is mostly concerned with the phenomenological signatures of this class of models, yet it is useful to review how the different mass scales are motivated, see Fig.~\ref{fig:model}. In the context of asymmetric dark matter, it is usually assumed that some mechanism relates the dark matter asymmetry to the baryon asymmetry. Since the observed dark matter energy density is about five times larger than the baryonic energy density, the dark matter mass should be of order $5\times m_{\rm proton}$, up to order one factors that depend on the exact mechanism of asymmetry sharing.\footnote{In the literature one can also find models where the ratio of number densities can vary over a larger range~(e.g.~\cite{Buckley:2010ui,Falkowski:2011xh}), in which case the motivation for GeV-scale dark matter is lost.}
In our case, the dark baryon is the dark matter candidate and has a mass of order $\Lambda_d$ giving the main motivation for considering $\Lambda_d$ in the $(1\, - \, 10)$ GeV range.  A dynamical mechanism to relate the dark confinement scale $\Lambda_d$ to the QCD scale was presented in~\cite{Bai:2013xga}, and other possibilities to motivate the GeV scale for dark matter can be found e.g. in~\cite{Nussinov:1985xr,Kaplan:1991ah,Barr:1990ca,Barr:1991qn,Dodelson:1991iv,Fujii:2002aj,Kitano:2004sv,Farrar:2005zd,Gudnason:2006ug,Kitano:2008tk,Kaplan:2009ag,Shelton:2010ta,Davoudiasl:2010am,Buckley:2010ui,Blennow:2010qp,Cohen:2010kn,Frandsen:2011kt}. 

A mediator that communicates between the dark and visible sectors is, in general, required for implementing a mechanism that shares the asymmetry and to allow an efficient annihilation of the symmetric relic density back to SM particles. In models with QCD like composite DM, the annihilation of dark baryons with dark anti-baryons into dark pions is typically very efficient so the dark baryon relic density is determined by the dark matter asymmetry. Entropy transfer back to the visible sector then happens via decays of dark pions. In order to not interfere with Big Bang Nucleosynthesis (BBN), the dark pion lifetime should be shorter than about one second, which implies a rather loose upper bound on the mediator mass of the order of 100~TeV. 
In~\cite{Bai:2013xga}, bifundamental mediators ensure a specific ratio of the QCD and dark QCD gauge couplings at the mass scale $M_{X_d}$. It was shown there that lower mediator masses are more likely to lead to a dark QCD confinement scale close to the QCD scale, such that within this model TeV scale mediators are preferred.

\subsection{Dark Pions}
\label{sec:darkpions}

As mentioned above, the lightest composite states are the dark pions $\pi_d$ which are the Goldstone bosons of the $n_f\times n_f$ dark flavor symmetry. The couplings Eq.~(\ref{eqn:darkYukawa}) break the global flavor symmetry such that small masses for the pions will be generated. 
Integrating out the heavy $X_d$ fields leads to an effective Lagrangian for the dark quarks of the form
\begin{align}
	m_{ij} \bar{Q}_{Li} Q_{Rj} + \kappa_{i \alpha} \kappa^*_{j \beta} \frac{1}{M_X^2} \bar{Q}_{Li} \gamma_\mu Q_{Lj} \, \bar{d}_{R\alpha} \gamma^\mu d_{R\beta} + {\rm h.c.} \,.
	\label{eqn:effOp}
\end{align}
Here one has to keep in mind that the explicit Dirac mass terms  $m_{ij}$ are not necessarily aligned in flavor space with the Yukawa couplings $\kappa$. The same effective Lagrangian would also arise from integrating out a $Z_d$ mediator. 

We now estimate the dark pion lifetime following the results of~\cite{Strassler:2006im,Bai:2013xga}. The lifetime can be quite suppressed relative to the naive order of magnitude estimate of $\Gamma \sim \kappa^4 m_{\pi_d}^5/(32 \pi M_{X_d}^4)$, depending on the structure of $\kappa$ and the masses of the dark pions. The dark quark current $j_\mu^D = \bar{d}_{Ri} \gamma^\mu d_{Rj} $ matches onto a dark pion current of the form $f_{\pi_d} \partial_\mu \pi_{d ij}$, where $f_{\pi_d}$ is the dark pion decay constant. Assuming universal masses and couplings for all dark pions, and assuming that $m_{\pi_d} > \Lambda_{\rm QCD}$, we obtain the decay width of dark pions into pairs of down-type quarks as~\cite{Bai:2013xga}
\begin{align}\label{eqn:piwidth}
	\Gamma(\pi_d \to \bar{d} d) & = \frac{\kappa^4 N_c  f_{\pi_d}^2 m_{\rm down}^2 }{32 \pi M_{X_d}^4 } m_{\pi_d}\,.
\end{align}
Here $N_c$ is a Standard Model color factor and $m_{\rm down}$ denotes a SM down type quark mass which arises from the chirality flip required for a pseudoscalar to decay to two fermions. We can now compute the proper lifetime:
\begin{align}
	c \tau_0 & = \frac{c \hbar}{\Gamma} \approx 80 \,{\rm mm} \times \frac{1}{\kappa^4} \times 
	\left( \frac{2~{\rm GeV}}{f_{\pi_d}} \right)^2 
	\left( \frac{100~{\rm MeV}}{m_{\rm down}} \right)^2
	\left( \frac{2~{\rm GeV}}{m_{\pi_d}} \right)
	\left( \frac{M_{X_d}}{1~{\rm TeV}} \right)^4\,.
\label{eq:lifetime}
\end{align}
It is therefore well motivated to consider centimeter to meter decay lengths for GeV scale dark pions with TeV scale mediators. There is some implicit sensitivity to the kaon threshold: when decays to kaon pairs are kinematically forbidden, the lifetime will increase by a factor of 400 and the dark pions tend be long lived enough to escape the detector unless the mediator mass is lowered. 

One can also imagine different electroweak quantum numbers for the bifundamental such that decays to up-type quarks are allowed. In this regime, decays to charm quarks would tend to dominate if kinematically allowed. Because charm hadrons have their own finite lifetimes, the decay of a dark pion could be a multi-stage process with the dark pion flying a finite distance and then decaying to charm hadrons which themselves travel through the detector before decaying to lighter states. This sort of phenomenology could also occur in the more extreme regions of parameter space where dark pion decay to $b$-quarks is kinematically accessible. The search strategies presented in the subsequent sections of this work will still be effective in the case of these heavier flavor decays. See App.~\ref{sec:dark_sector} for further discussion. 

Eq.~\ref{eq:lifetime} is the origin for the 100~TeV bound on the mediator mass - for higher mediator masses the dark pion lifetime will get dangerously close to the BBN time. Apart from this bound, the dark pion properties are of minor importance for the cosmology of this model. On the other hand, the collider phenomenology will be dominated by meson production, with the dark baryon multiplicity being much smaller for QCD like theories~\cite{Beringer:1900zz}, and even further suppressed in the large $N_d$ limit~\cite{Witten:1979kh}. Since one can expect that all heavier dark mesons decay to dark pions on a time scale given by $\Lambda_d^{-1} \ll \Gamma (\pi_d \to \bar{d} d)^{-1}$, the dark pion lifetime will be crucial to determine where the dark jets will emerge in the detector.

%
%%%%%%%%%%%%%%%%%%%%%%%%%%%
\section{Emerging Jet Phenomenology}
\label{sec:ph}
%%%%%%%%%%%%%%%%%%%%%%%%%%%
%

\subsection{Collider Signal}
At a hadron collider, the mediator particles can be produced on-shell provided that their mass is sufficiently below the center-of-mass energy of the experiment. Here and in the following we will mostly focus on the production of $X_d\, X^\dagger_d$ pairs through a virtual gluon, which can be initiated both from quark and gluon initial states.

\begin{figure}
\center
\includegraphics[width=0.4\textwidth]{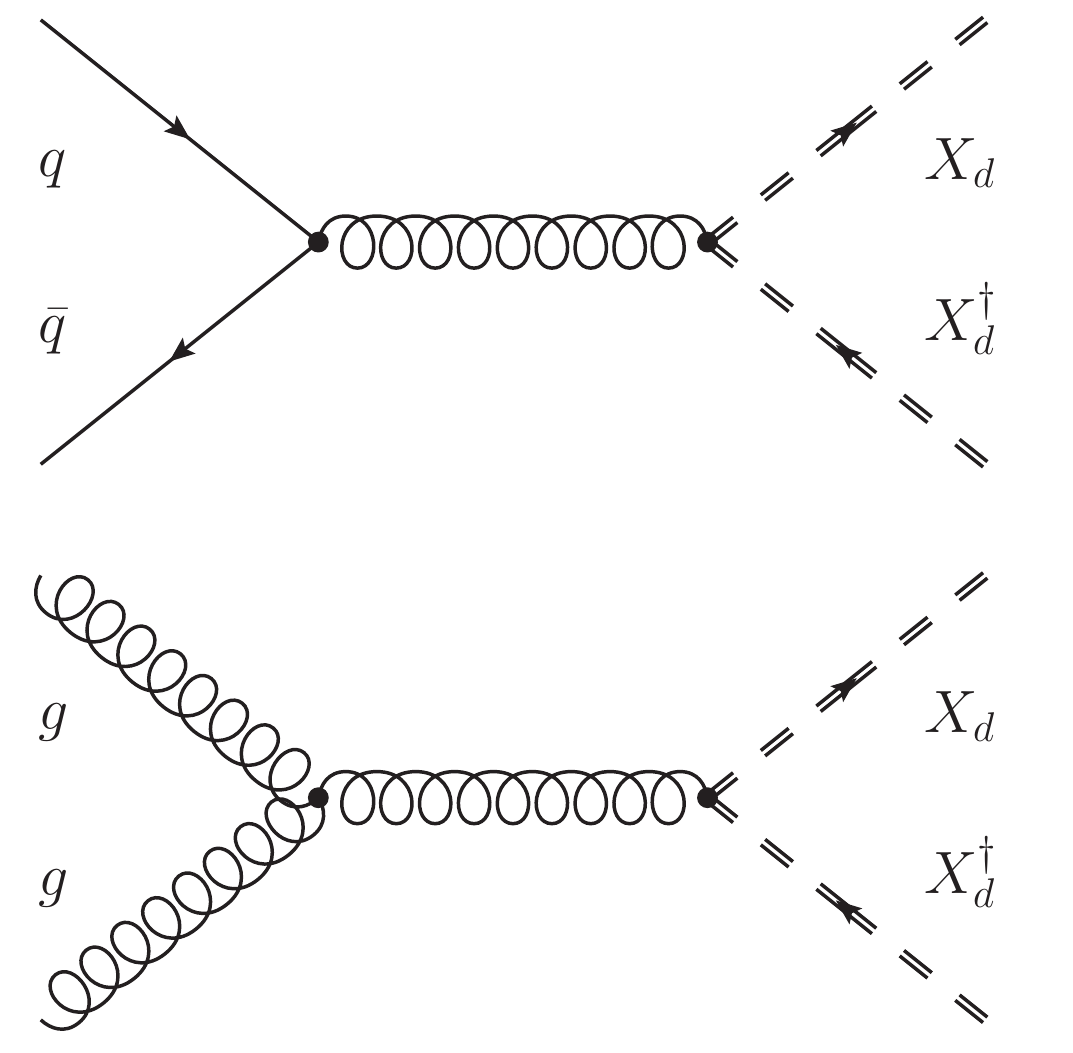}
\hspace*{.2cm}
\includegraphics[width=0.55\textwidth]{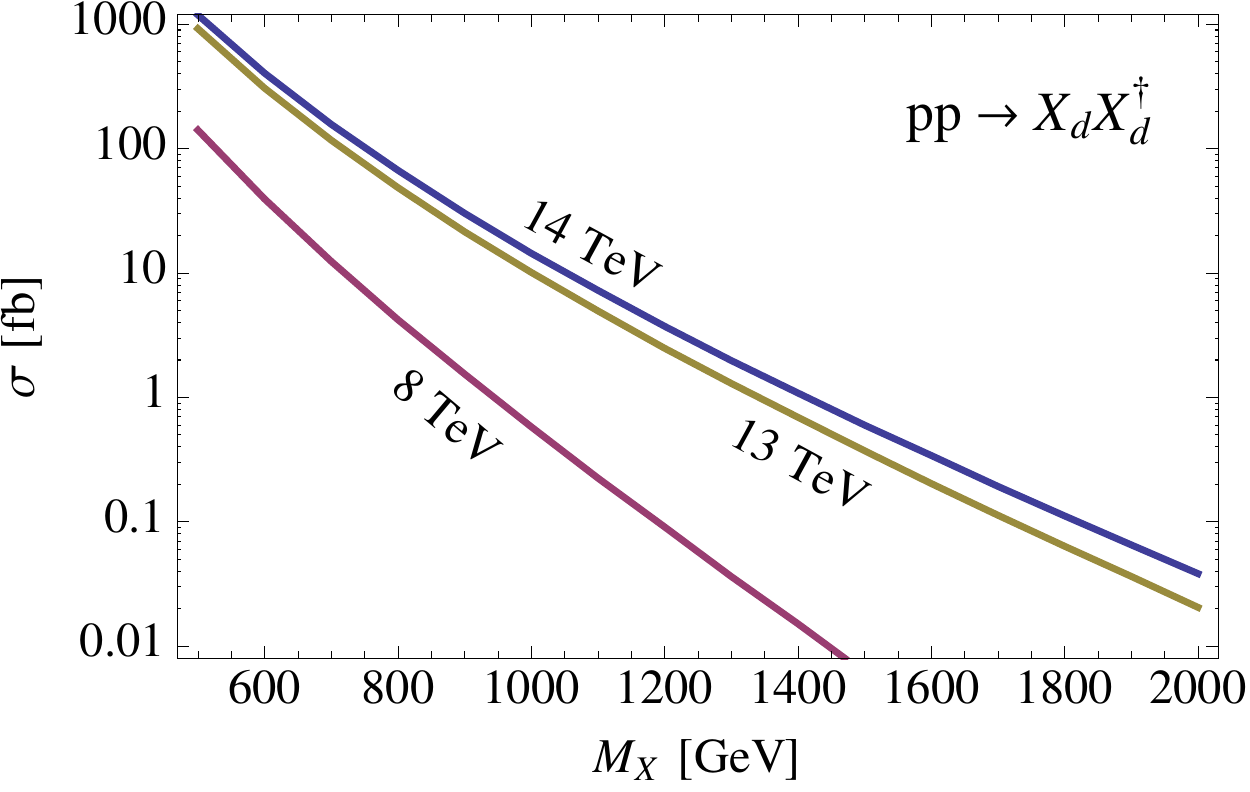}
\caption{Left: Feynman diagrams for the pair production of $X_d$ at hadron colliders. 
Right: Tree level cross section for $X_d$ pair production at the LHC.  }
\label{fig:xsec}
\end{figure}

The most important diagrams that contribute to the production are shown in Fig.~\ref{fig:xsec}. Apart from the dark color degrees of freedom, the production process is very similar to pair production of one squark flavor in supersymmetry and is set by QCD gauge invariance. Therefore the cross section is similar, for example, to that of pair production stop quarks multiplied by $N_d$. In the plot on the right of Fig.~\ref{fig:xsec} we show the tree level cross section for $X_d \,X^\dagger_d$ production for different center of mass energies at the LHC, obtained from \textsc{Pythia}\footnote{Throughout this work, we use a modified version of \textsc{Pythia} 8.183, see \url{https://github.com/pedroschwaller/EmergingJets}, and we use the default tune 4C unless otherwise specified.}~\cite{Sjostrand:2007gs} using \textsc{CTEQ 6.1} parton distribution functions (PDF)~\cite{Pumplin:2002vw}. Since the parton luminosity for quark-gluon initial states is large at the LHC, next-to-leading order corrections that include the process $p p \to X_d \, X^\dagger_d \, j$ can be sizable. Based on the similarity with squark production with decoupled gluinos, we can expect a K-factor of around 1.3~\cite{Beenakker:1996ch}, but we use tree-level cross sections for our subsequent analyses.

If the mediator $X_d$ has order one couplings  $\kappa_{ij}$ to the quarks and dark quarks, it will decay before the onset of hadronization both in QCD and dark QCD. Therefore we can treat the pair production of $X_d$ with subsequent decay $X_d \to Q_d \bar{q}$ as hard process. The SM quarks from $X_d$ decays will produce ordinary QCD jets. On the other hand, each dark quark $Q_d$ will first undergo parton showering and fragmentation in the dark sector, which happens on a time scale $\Lambda_d^{-1}$, much shorter than the time scale for dark mesons to decay back to SM particles. 

In order to explore the resulting phenomenology, we should therefore first understand the structure and basic features of the dark parton shower and fragmentation. The dark parton shower, i.e. the radiation of dark gluons off dark partons, and the splitting of dark gluons into dark quark pairs, in non-abelian gauge theories is theoretically well understood and described by so called DGLAP~\cite{Gribov:1972ri,Altarelli:1977zs,Dokshitzer:1977sg} evolution equations. It essentially depends on the running of the coupling, i.e. on the number of colors and quark flavors. Unless the theory is in the conformal window, the jet objects should be similar to QCD jets. 

Fragmentation, the conversion of dark partons into dark hadrons, is a non-perturbative process that can only be modeled even for QCD, so we have to infer from QCD for the dark sector. As discussed above, the production of baryons is suppressed relative to meson production in the large $N_c$ limit~\cite{Witten:1979kh}, and happens at the 10\% level in QCD~\cite{Beringer:1900zz}. Among the dark mesons the most important distinction is between Goldstone bosons $\pi_d$, with masses below $\Lambda_d$, and heavier resonances with masses of order $\Lambda_d$. The latter ones will decay to the lightest available states (i.e. the Goldstones) on very short time scales of $1/\Lambda_d$. Therefore, when a dark quark is produced at a collider, it undergoes showering and then hadronization into a jet composed mostly of dark pions, $\pi_d$, originating from the interaction point (but invisible to the detector before they decay). The typical dark jet will have a small fraction of its energy in dark baryons that escape the detector and give rise to some missing energy, but given the large uncertainties on jet energy measurements, this is will be an unimportant effect for most jets.\footnote{For models where the average jet will have a larger fraction of missing energy, a search strategy was presented in~\cite{Cohen:2015toa}.} 

The ``dark jet'' production is shown schematically in Fig.~\ref{fig:emerging-jets}, with the dark pions represented by grey dashed lines. Depending on their lifetime, the dark pions may travel a measurable distance away from the interaction point before decaying to SM particles. In the laboratory frame, the characteristic decay length is given by $ \beta\, \gamma \,c\,\tau_{\pi_d}$, where $\beta \gamma$ is the boost factor that depends on the momentum of each individual pion. Furthermore since the actual decay time is distributed exponentially, each pion will decay at a different distance from the interaction point, with harder particles traveling further on average. 

\begin{figure}
\centering
\includegraphics[width=0.4\textwidth,valign=t]{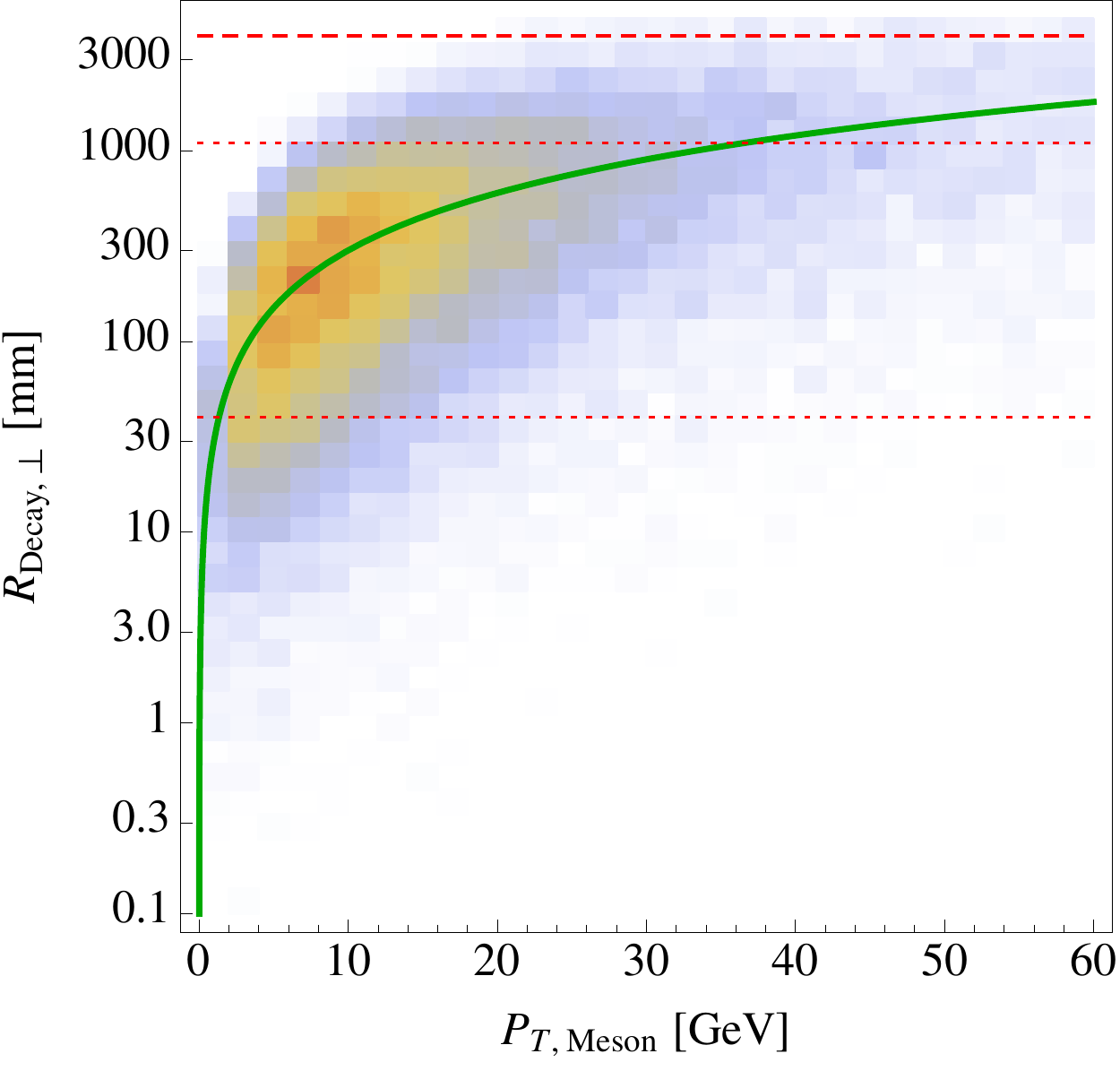}\hspace{.5cm}
\includegraphics[width=0.4\textwidth,valign=t]{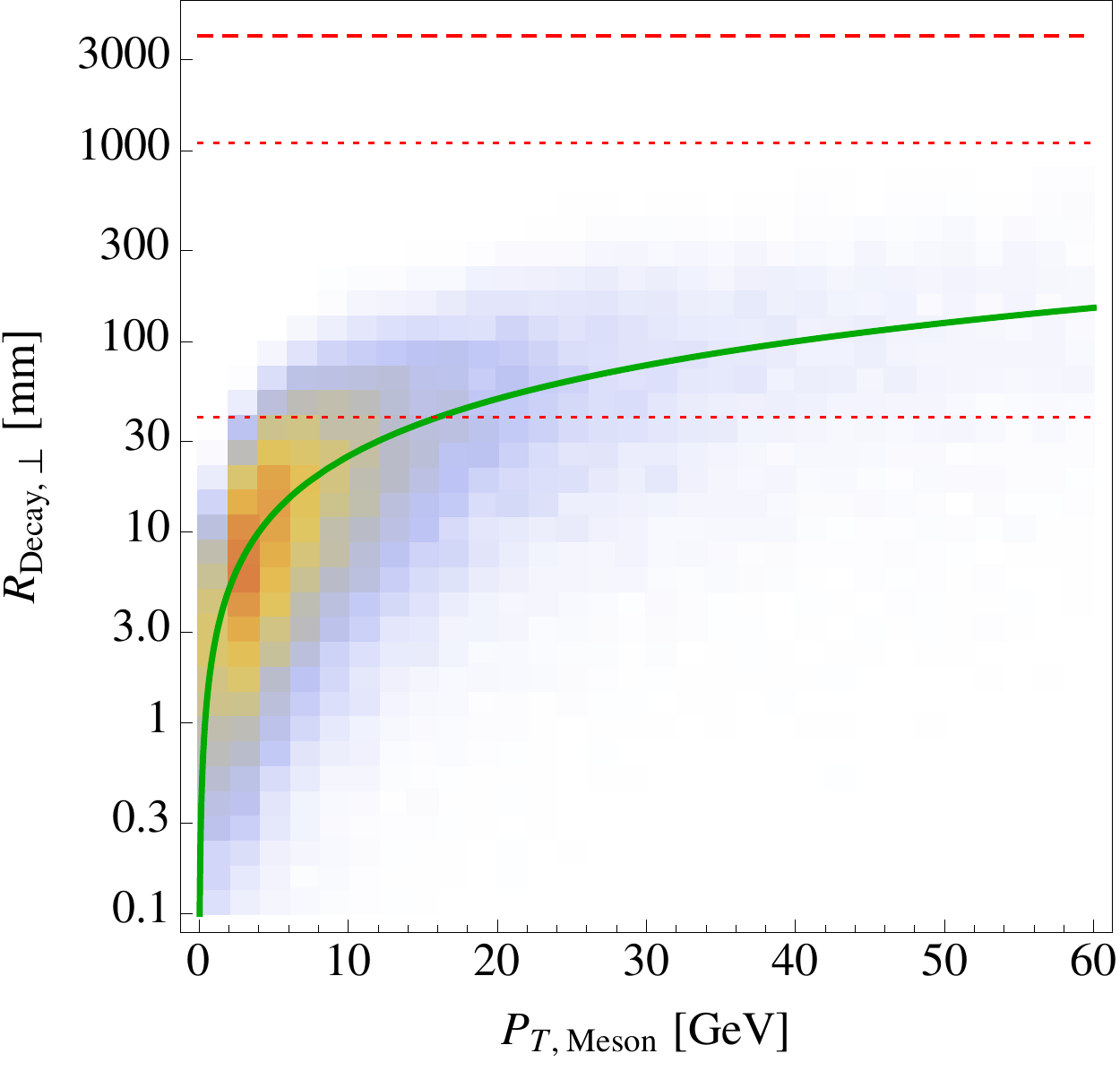}\hspace{.5cm}
\includegraphics[width=0.06\textwidth,valign=t]{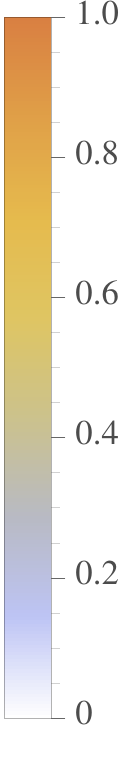}
\caption{Distribution of transverse decay distances of individual dark pions for model A (left) and model B (right) at LHC14 (the benchmarks are defined in Sec.~\ref{sec:benchmarks}). The green curve shows the average transverse laboratory frame decay length $\beta_T \gamma_T c\tau_{\pi_d} = (p_T/m_{\pi_d}) c \tau_{\pi_d}$. Dashed lines indicate the approximate regions covered by the tracker (50~mm - 1000~mm) and calorimeters (1000~mm - 3000~mm).  }
\label{fig:decaydist}
\end{figure}

In order to simulate production and dynamics of the dark sector at the LHC, we use a modified version of the Hidden Valley implementation~\cite{Carloni:2010tw,Carloni:2011kk} of \textsc{Pythia}~\cite{Sjostrand:2007gs}, and we describe the details of the simulation in App.~\ref{sec:simulation}. Armed with this simulation and our benchmarks described in Sec.~\ref{sec:benchmarks}, we can begin a quantitative study of the dark sector.  In Fig.~\ref{fig:decaydist} we show the distribution of transverse decay distances from the interaction point for two benchmark models, see Sec.~\ref{sec:benchmarks} for their definition. The majority of decays occurs well away from the beam pipe, but still within the tracker, and are clustered around the average transverse decay length $\beta_T \gamma_T c\tau_{\pi_d} = p_T/m_{\pi_d} c \tau_{\pi_d}$. From here we can easily understand what a change of parameters will imply: the average decay distance will change proportional to the proper lifetime and inversely proportional to the mass of the dark pions for fixed mediator mass. Given the physical size of the trackers and hadronic calorimeters, we can easily vary the parameters by one to two orders of magnitude without changing the signal in a significant way. We further explore what happens when different parameters are varied in App.~\ref{sec:dark_sector}.

Before the dark pions decay, the jet is completely invisible, so we now describe this decay back into the visible sector. 
When the dark pion decays to SM quarks, it will produce a sub-jet with a small number of SM hadrons all originating from a common displaced vertex. This is depicted by the solid colored lines in Fig.~\ref{fig:emerging-jets}. The average multiplicity of the sub-jets will depend on the dark pion mass. As we will see below, LHC searches exist which are optimized to search for a single displaced vertex, but there is no search which looks for many nearby vertices. If we examine the jet at a distance which is large compared to the typical $\gamma \, \beta\, c\,\tau_{\pi_d}$, we see many SM hadrons going in the same direction: an object that very much resembles a standard jet. Therefore, if using only calorimeter information, the usual techniques that measure jets will work well. On the other hand, if we look at the radial profile of the jets, we see that at the interaction point there is very little visible energy, and there is more and more as one is further from the initial interaction point. The jet \textit{emerges} within the detector, producing a very distinct signature.\footnote{It should also be noted that this signature is distinct from the "trackless jets" considered in~\cite{Bai:2011wy}, which have absolutely no tracks and also potentially non-standard interactions with the calorimeter.}

\subsection{Existing Searches and Constraints}

In the following section, we will present a detailed search strategy for such \textit{emerging jets} at the LHC, but we will first discuss existing searches for displaced objects and why they are not sensitive to emerging jets. 

First, pair production of $X_d$ produces a 4-jet signature at the calorimeter level, with pairs of jets reconstructing the $X_d$ mass. Searches in this channel have been performed by the ATLAS~\cite{ATLAS:2012ds} and CMS~\cite{Chatrchyan:2013izb,Khachatryan:2014lpa} experiments and have been interpreted in terms of RPV stop decays. Taking into account the $N_d$ enhancement of $X_d$ pair production compared to MSSM stops, the most recent CMS results~\cite{Khachatryan:2014lpa} would imply a limit of $M_{X_d} \gtrsim 600$~GeV. This interpretation is not straightforward however. The CMS search utilizes jets reconstructed using a particle flow algorithm, which includes tracking information, and the sensitivity was estimated assuming prompt jets. Furthermore there is a possibility that jet quality cuts will remove some or all of the emerging jets. Therefore values of $M_{X_d}$ lower than 600~GeV can not be conclusively excluded from that search alone. 

Effects of new colored states can also be probed indirectly, for example through their effect on the running of the strong coupling constant. The most recent measurement of $\alpha_s(Q)$~\cite{CMS:2014mna} shows no deviation in the strong coupling up to $Q\approx 1.4$~TeV, but is not yet sensitive enough to exclude additional colored states above the weak scale. Furthermore the mediators $X_d$ could contribute to the dijet cross section, if the emerging jets would be reconstructed as ordinary jets. In that case one would obtain a bound on the couplings of $X_d$ to first generation quarks, which depends on the flavor structure of the model, but not directly on the mass. 

Apart from generic multi-jet searches, several analyses dedicated to displaced or otherwise exotic jet signatures exist. 

\begin{figure}
\centering
\includegraphics[width=0.7\textwidth]{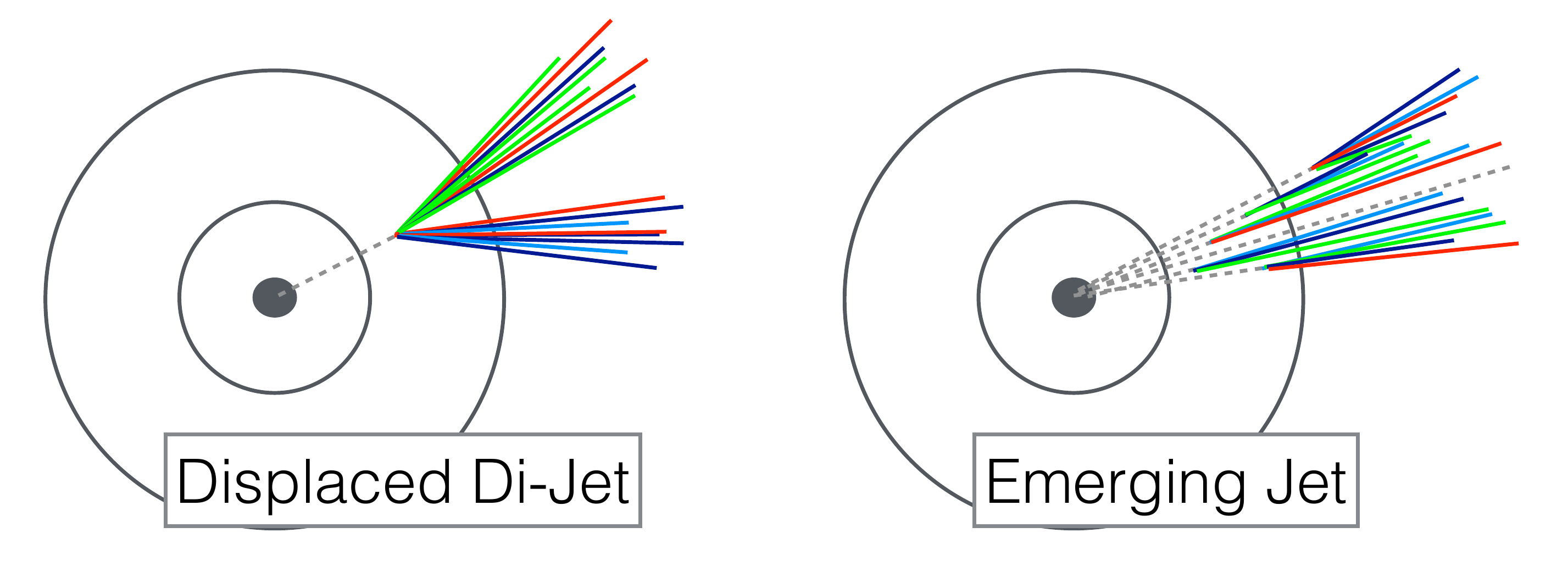}
\caption{
Difference between a displaced dijet signature from the decay of a heavy long-lived particle and the emerging jet signature. 
}
\label{fig:dijet}
\end{figure}

{\bf{CMS displaced dijet search}}: CMS has a search for pair production of a long lived particle which decays to two jets~\cite{CMS:2014wda}. Two distinct jets with $p_T > 60$~GeV and a separation of $\Delta R > 0.5$ are required and are fitted to the same displaced vertex. This differs qualitatively from the emerging jets scenario as shown in Fig.~\ref{fig:dijet}, and this can be seen from the specific analysis strategy employed in~\cite{CMS:2014wda}. 
In order to reduce background from pile up, this search requires one good vertex with at least 4 GeV invariant mass and 8 GeV $p_T$. Once that vertex is constructed, it eliminates tracks which do not pass through that vertex. Most emerging jet events will already fail the requirement of having two displaced jets that originate from the same vertex, as illustrated in Fig~\ref{fig:dijet}.
Furthermore, in the emerging jet scenario with many different displaced vertices, this algorithm will have difficulty choosing a vertex and then will throw out the majority of the tracks, drastically reducing the signal efficiency. While this search is difficult to accurately recast, it is clearly not optimal, and it is unlikely to be sensitive to the emerging jet signal.

{\bf{ATLAS displaced event triggers}}: ATLAS has published a description of triggers~\cite{Aad:2013txa} that can be used for displaced events. As we will see below, triggering is not a problem for our signal because of the energy deposited in the calorimeters. The main ATLAS trigger for objects that decay before reaching the calorimeter requires zero tracks reconstructed using the standard algorithm within the jet cone. It also requires a muon inside that cone with $p_T > 10$ GeV, and neither of these requirements are generic in emerging jet scenarios. There are also triggers for long-lived particles decaying in the calorimeters or muon system, but we do not focus on that region of parameter space here. 

{\bf{ATLAS long lived neutral particle search}}: ATLAS has also published a search of long lived neutral particles~\cite{Aad:2015asa} and one for lepton jets~\cite{Aad:2014yea}. In our case, we generically have pair production of a long lived object which then decays to two or four states, so as with the CMS search, the models considered only has one displaced vertex for each exotic object. Both searches require the EM fraction, the fraction of energy in the electromagnetic calorimeter relative to the hadronic calorimeter, to be smaller than 0.1.\footnote{The lepton jet search only requires this for their hadronic category, but the categories that require muons will also not be sensitive.} This requirement is designed to select objects decaying in the hadronic calorimeter and thus leaving very little energy in the electromagnetic one. Because of the emerging nature of the signal considered here, there will be energy in all segments of the calorimeter and this cut would generally cut out the majority of our signal. It could be sensitive to regions of parameter space with longer lifetimes, but then there will be quite a few dark decays in the muon system and it is not clear how they will be reconstructed. In the region of parameter space we are most interested in, the EM fraction cut will make the signal efficiency extremely low for emerging jets. 

{\bf{LHCb displaced dijet search}}: LHCb has a search~\cite{Aaij:2014nma} which is based on a similar model as the aforementioned CMS search. They also require reconstruction of a single vertex and force the majority of particles to pass through (or near to catch $b$ and $c$ hadrons) this vertex. Therefore, if there are many hard vertices displaced from one another by a few millimeters then this search will have low efficiency for the emerging phenomenology considered. Because of the relatively small geometric acceptance, there will be events where only one dark pion falls into LHCb, and the analysis could be sensitive in this regime. All the limits described in the analysis, however, are for dark pion mass above 25 GeV, so it is a somewhat different regime of the model than we consider. More details will be given about the LHCb potential in Sec.~\ref{sec:lhcb}. It should also be noted that the searches discussed above constrain models with mediators in the 100~GeV range and with pico barn cross sections, while we are aiming at TeV scale mediators. 

{\bf{Other long-lived particle searches}}: The remaining published searches for long lived particles and/or displaced decay topologies often require additional isolated leptons (e.g.~\cite{Aad:2011zb,TheATLAScollaboration:2013yia,CMS:2014mca}) or use timing information (e.g.~\cite{Chatrchyan:2012dxa,Aad:2013gva}) to distinguish from SM backgrounds. The emerging jet signature discussed here possesses neither of these features. Therefore, in the next section we present a potential search strategy for discovery at the LHC.

%--------------------------------------------------------------------%
\section{Analysis Strategy}
\label{sec:analysis}
%--------------------------------------------------------------------%

Here we present our analysis strategy to search for dark sectors with bi-fundamental mediators. The tracking system in both the ATLAS and CMS detectors extends from about $50$~mm to $1$~m from the interaction point in radial direction. Tracks can be reconstructed with a resolution of about 100~$\mu$m in the impact parameter for charged pions with $p_T > 1$~GeV, and the track reconstruction efficiency is above $95\%$ for central pions and above 90\% in the forward region~\cite{Chatrchyan:2014fea}. 

While the tracker starts a few centimeters from the beamline, there are several possibilities to determine whether a track originates from the primary vertex with a precision as small as a few hundred micrometers. First, the impact parameter itself can be used to determine whether a track originates from the primary vertex. A more powerful technique that is usually employed by the experiments is to reconstruct secondary vertices and to measure their transverse distance $L_{xy}$ from the primary vertex (see e.g.~\cite{CMS:2014wda}). In the following we will assume that this technique can be employed to determine the trackless distance of a jet object down to at at least a millimeter. After presenting the general analysis strategy, we will discuss this in more detail in Sec~\ref{sec:emergingjets}, and the details of how we simulate detector response are given in App.~\ref{app:det-sim}.

\subsection{Benchmarks}
\label{sec:benchmarks}

In this section we will describe some of the parameters of the dark sector and the mediator, and we will define the benchmark models that we will analyze in the rest of the paper.  We take our benchmark value for the mediator mass $M_X$ to be 1 TeV, though we will vary this parameter in order to estimate the LHC reach for these scenarios. For the dark sector parameters, we consider two benchmark parameter points that capture the relevant phenomenology and allow us to study which observables are model dependent and which are relatively robust within this framework. The benchmark points are shown in Tab.~\ref{tab:benchmarks}. Inspired by QCD, we take the dark vector masses to be somewhat heavier than the confinement scale $\Lambda_d$, and we take the dark pion masses to be lighter for both benchmarks. This means that dark vectors will undergo rapid decay into dark pions before they can decay into SM hadrons. 

\begin{table}[tb]
\centering%%
\begin{tabular}{|c|c|c|}
\hline
 & Model \textbf{A} & Model \textbf{B}  \\ \hline 
$\Lambda_d$ & 10 GeV  & 4 GeV \\
$m_V$ & 20 GeV  & 8 GeV  \\
$m_{\pi_d}$ & 5 GeV  & 2 GeV \\
$c\,\tau_{\pi_d}$  & 150 mm  & 5 mm \\ \hline
\end{tabular}
\caption{Dark sector parameters in our two benchmark models. $\Lambda_d$ is the dark confinement scale, $m_V$ is the mass of the dark vector mesons, and $m_{\pi_d}$ is the pseudo-scalar mass. $c\,\tau_{\pi_d}$ is the rest frame decay length of the pseudo-scalars. We take $N_c = 3$ and $n_f = 7$ in both benchmarks. }
\label{tab:benchmarks}
\end{table}

Model A describes a somewhat heavier dark sector such that an average of $\mathcal{O}(10)$ visible hadrons will be formed in each dark pion decay, while model B is lighter and there will only be a few visible hadrons per dark pion decay (particle multiplicity will be discussed in greater detail in Sec.~\ref{sec:lhcb}). Model A also has a relatively longer lifetime so that a substantial fraction of the dark meson decays will occur in the calorimeters or beyond, while model B has a short lifetime and most decays occur within the tracker.  In App.~\ref{sec:dark_sector} we further explore the parameter space of the dark sector and describe how our analysis is relatively robust throughout. We also give examples of collider level observables that are sensitive to the dark sector parameters. The search strategy that we will present in the following is largely independent of the details of the dark sector.

\subsection{Triggering}

Pair production of the mediators $X_d$ leads to four calorimeter jets, so we propose to trigger on four or more hard, central jets. Such triggers were employed for example in the paired dijet resonance search by CMS~\cite{Chatrchyan:2013izb,KaiYifortheCMS:2013vqa} and in a search for pair production of massive colored scalars by ATLAS~\cite{ATLAS:2012ds}.  The CMS search requires at least four jets with $p_{T,j}>80$~GeV and $|\eta|<3.0$, based on calorimeter information, and the trigger is 99.5\% efficient for events with $p_{T,j}>110$~GeV and $|\eta|<2.5$ for each jet. It should be noted that while CMS ultimately relies on particle flow jets for the analysis, the triggers only utilize calorimeter information. Similarly in the  ATLAS search a four (or more) jet trigger is used with is 99\% efficient for $p_{T,j}>80$~GeV. 

For the 13/14~TeV run of the LHC, the trigger thresholds will most likely increase. We will use jets with $p_T>200$~GeV and $|\eta|<2.5$ for the analysis, which should be well above the minimum trigger requirements of the upcoming LHC run. 

Since the triggers are based on calorimeter information, the emerging jet properties do not pose a problem at this stage. On the other hand certain jet quality requirements could lead to the events being discarded. The two jets originating from SM quarks guarantee a well reconstructed primary vertex for the hard process, and will allow efficient rejection of pile-up. The emerging jets will have tracks pointing towards the calorimeter energy deposits that do not originate from the primary vertex. It will be important to make sure that jet reconstruction algorithms that utilize tracking information do not reject those jets as calorimeter noise or other non-collision background. The simplest possibility here would be to use pure calorimeter jets for this analysis. On the other hand, since there will be emerging tracks, it should be possible to utilize more advanced jet reconstruction techniques, provided that they are flexible enough to not reject emerging jets. 

\subsection{Event Selection}
\label{sec:events}

We now analyze the $X_d$ model at LHC14.  The typical signal event has two emerging jets and two standard QCD jets, so this search is similar to current LHC searches for paired dijet resonances~\cite{ATLAS:2012ds,Chatrchyan:2013izb}, and our cuts are loosely inspired by these searches.  We cluster the jets using the FastJet~\cite{Cacciari:2011ma} implementation of the anti-$k_t$ algorithm~\cite{Cacciari:2008gp} with $R = 0.5$. We demand at least four jets with $p_T > 200$ GeV and $|\eta|<2.5$, and we also require that the scalar sum of the $p_T$'s of those jets is greater than 1000 GeV. The efficiency of the kinematic cuts is 34\% (58\%) for model A (B) with an $X_d$ mass of 1 TeV that we take for the rest of this section. The cut flow for this analysis is shown in Table~\ref{tab:cut-flow4}. The experimental searches for paired resonances~\cite{ATLAS:2012ds,Chatrchyan:2013izb} also cut on the difference between the dijet invariant masses, which gives a moderate improvement in signal to background, but we do not use it here because the emerging jet cut described below will be so effective. 

\begin{figure}
\centering
\includegraphics[width=\textwidth]{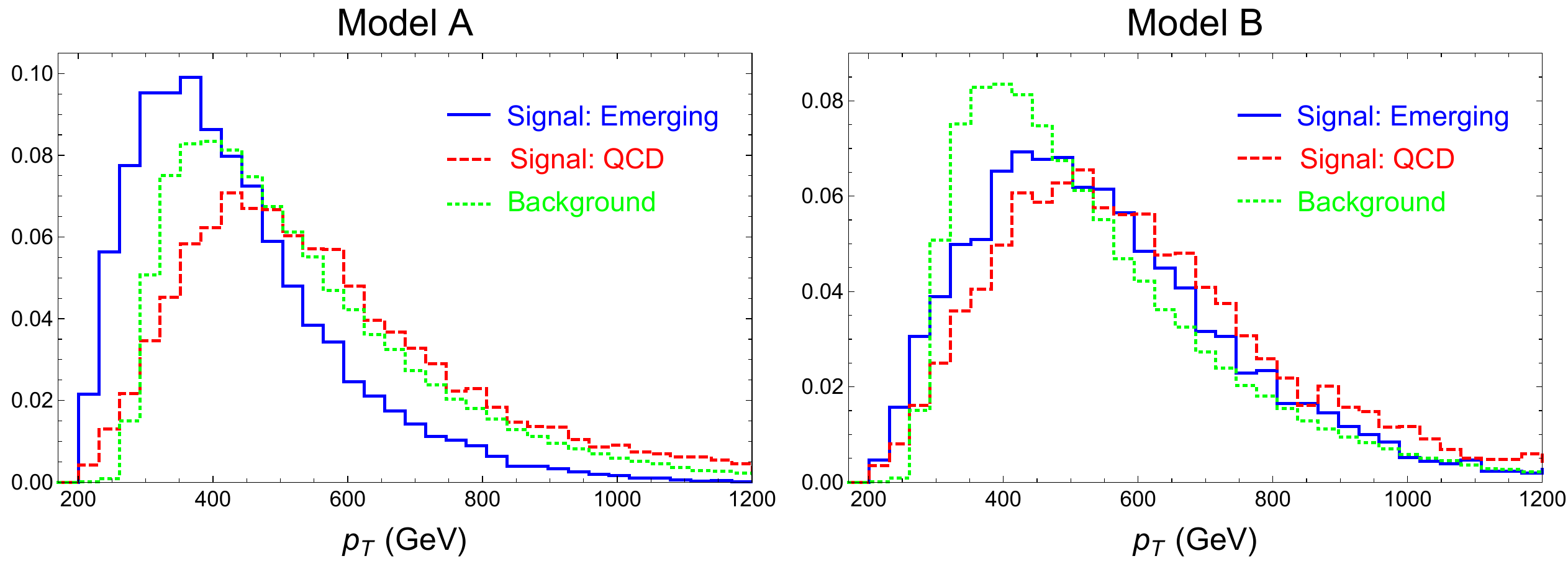}
\caption{$p_T$ distributions of the hardest emerging (solid, blue) and hardest QCD (dashed, red) jet in each signal event, as well as for the hardest jet in the background QCD sample (dotted, green). Emerging jets have $r=3$ mm, $n=0$, and $p_T^{\rm min} = 1$ GeV. These events pass all the kinematic cuts described in the text, and the signal events have at least two emerging jets. The left plot is for model A, while the right for model B.  }
\label{fig:pt}
\end{figure}

It is important to know the kinematic features of our signal events. In Fig.~\ref{fig:pt} we plot the $p_T$ distribution for the leading emerging and non-emerging jet in each event that passes the kinematic cuts and has at least two emerging jets with $r=3$ mm and $n=0$ (see Sec.~\ref{sec:emergingjets} for details). We see that  these events tend to have quite hard jets with typical $p_T$ for the hardest jet $\mathcal{O}(500)$ GeV, which enables the trigger using multiple hard jets. 

From Fig.~\ref{fig:pt}, we also see that in model A the emerging jets tend to be softer than those from QCD. This is because we are taking jet energy as the energy deposited in the calorimeters (for details see App.~\ref{app:det-sim}), and in model A there will be many pions that decay beyond the calorimeters. These decays can in principle be measured by the muon systems of the experiments, but we leave the exploration of this feature to future work. The vast majority of model A events will have at least one meson decaying outside the calorimeter, with an average of three per event.  Furthermore, these mesons tend to carry a substantial amount of energy because they often have a large relativistic $\gamma$ factor. On the other hand, model B has a lifetime of 5 mm, so only 2\% of events have mesons that decay outside of the mock calorimeter. This explains why in Fig.~\ref{fig:pt} the $p_T$ distributions of emerging and non-emerging jets in model B are very similar.

\subsection{Backgrounds}

The dominant background for these sorts of four jet events will be from high $p_T$ QCD. We simulate four jet (including $b$) production in QCD using \textsc{MadGraph5\_aMC@NLO}~\cite{Alwall:2014hca} with \textsc{CTEQ 6.1} PDFs~\cite{Pumplin:2002vw} and hadronize using \textsc{Pythia}~\cite{Sjostrand:2007gs}. We apply parton level cuts that require each of the four jets to have $p_T > 150$ GeV and that the scalar sum of the $p_T$'s of the jets $H_T > 800$ GeV. This is the tree level cross section shown in Tab.~\ref{tab:cut-flow4} for the background. 

With just the kinematic cuts, the signal to background ratio is dauntingly small, $\mathcal{O}(10^{-4})$. Requiring emerging jets can dramatically reduce the background because the majority of QCD jets will have a large number of prompt tracks. QCD can fake the signal because the standard model has neutral hadrons with detector scale lifetimes such as the bottom and strange mesons and baryons. In addition, if we only insist on the absence of prompt tracks and not on the presence of displaced tracks, then QCD can produce jets dominated by long lived neutral hadrons (like the neutron) and photons.  As discussed in App.~\ref{app:det-sim}, we use a conservative photon rejection criteria, but the experiments can potentially do much better than we estimate at rejection photon dominated backgrounds. 

\begin{figure}
\centering
\includegraphics[width=1.\textwidth]{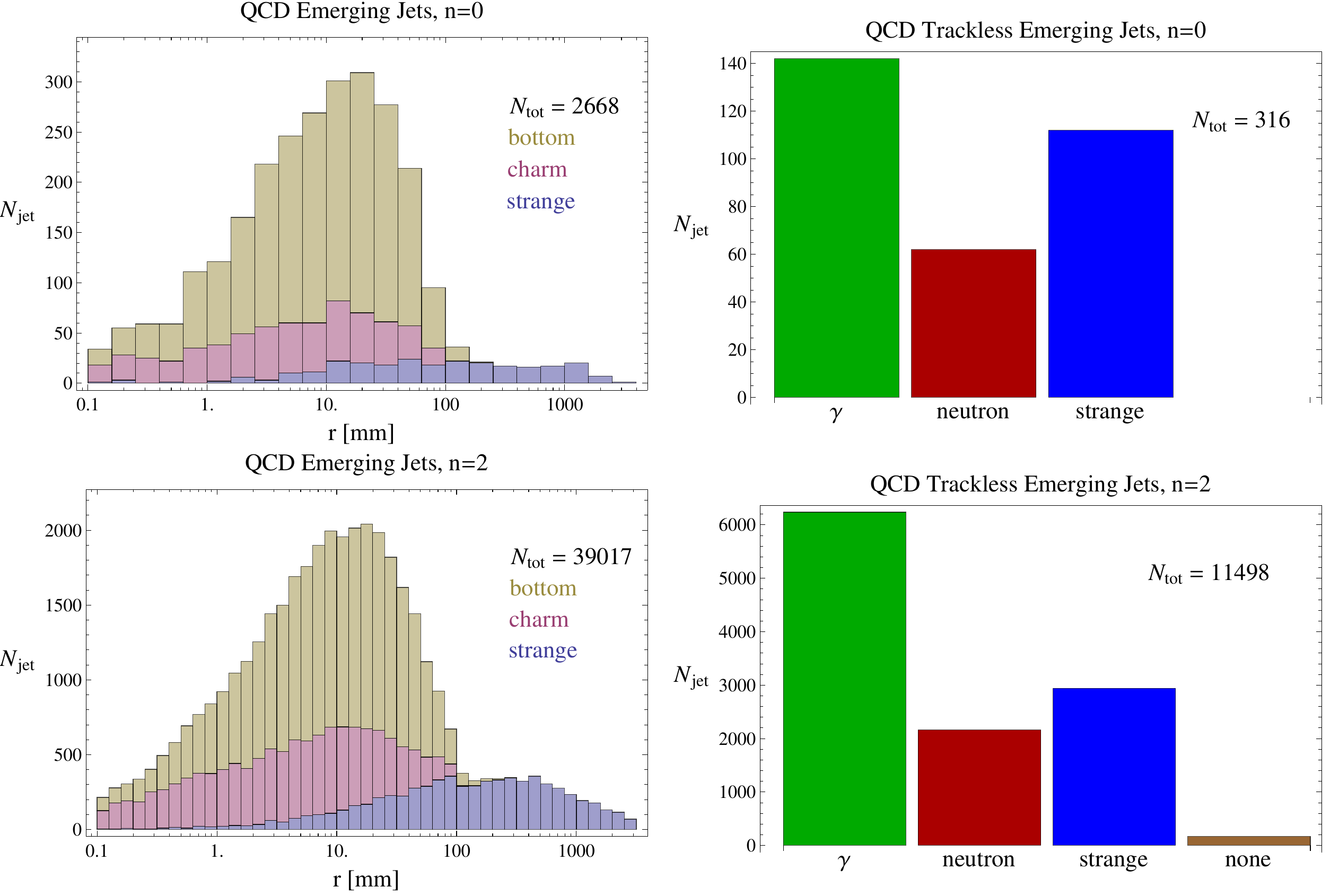}
\caption{Breakdown of the composition of the different ways that QCD can produce emerging jets. The left plots show the distribution of transverse decay radius of the earliest decaying neutral hadron within the jet. The histograms are stacked based on the quark content of the decaying neutral hadron, with strange, charm, and bottom going from bottom to top. The top (bottom) plot require $\leq 0$ (2) prompt charged tracks in the jet, and throughout we require all tracks to have $p_T > 1$ GeV. The right plots are jets with no displaced charged tracks at all and again $\leq 0$ (2) prompt charged tracks on the top (bottom). These jets are composed of photons, neutrons, neutral strange hadrons, and in the bottom plot, one or two prompt tracks. The right plots categorize these jets by which of the three types of displaced neutral categories carry the most $p_T$. The ``none'' category in the bottom plot is for jets where all the energy is in the one or two prompt tracks. All of the jets displayed must pass the kinematic cuts described in the text and in Tab.~\ref{tab:cut-flow4}. }
\label{fig:bgComp}
\end{figure}

In Fig.~\ref{fig:bgComp} we attempt to characterize the emerging jets produced within QCD. The plots on the left give the breakdown of jets which have at least one displaced track, and show where that track emerged and what type of neutral particle gave rise to it. The plots on the right describe jets with no displaced charged tracks at all. The top row requires that there are $n=0$ prompt tracks, while the bottom uses the looser requirement of $n\leq 2$ prompt tracks. We first note that requiring $n=0$, no prompt tracks, the background is dominated jets with some displaced tracks, while for $n=2$ the jets with no displaced tracks become a larger fraction. 
  
For jets with charged tracks, those with the earliest prompt track of transverse radius less than about 5 cm tend to be dominated by $b$-hadrons such as $B^0$ and $\Lambda_b$, while at larger radii, the sample is dominated by strange mesons and baryons such as $K_S^0$ and $\Lambda$. This figure was generated with $2\cdot 10^7$ QCD events, and, as described in App~\ref{app:det-sim}, this is for jets which deposit at least 200 GeV in the calorimeters. For jets with no charged tracks, we see that the energy of the jets is carried by either photons, neutrons, or strange hadrons, and all other species decay before reaching the calorimeters. A substantial fraction of trackless jets are dominated by photons which tend to come from $\pi^0$ decays. 

In addition to QCD backgrounds, there are also detector backgrounds which we do not attempt to simulate. These include interactions with the beam pipe or with other parts of the detector that can lead to displaced tracks. The nature and size of these backgrounds will vary greatly depending on the specific detector, therefore a full detector simulation is necessary to characterize them properly. On the other hand, non-collisional backgrounds are very unlikely to pass the stringent kinematic cuts we are imposing on the signal jets, therefore we do not expect them to qualitatively change our conclusions.

\subsection{Emerging Jets}
\label{sec:emergingjets}

We now come to the key cut in the analysis, the requirement that events contain emerging jets. We define an emerging jet as new reconstruction object $E(p_T^{\rm min}, n, r)$, to be a jet with $\leq n$ tracks with $p_T > p_T^{\rm min}$ originating a transverse distance smaller than $r$ from the interaction point. We can see this pictorially from Fig.~\ref{fig:emerging-jets} by drawing a circle around the interaction point and requiring that there are fewer than $n$ tracks above the $p_T$ threshold within that circle. The optimum size of the circle, $r$, will depend on the typical decay length of the dark pions. 

The innermost layers of the trackers at CMS and ATLAS are between about 50 and 100 mm from the $z$ axis, so for values of $r$ larger than roughly 100 mm, this strategy can be translated to looking for tracks that do not have any hits in the innermost layers of the tracker. For smaller values of $r$, there are two possibilities as to how to veto on tracks originating at a distance smaller than $r$. The first is the strategy employed in $b$-tagging, which is to look at the impact parameter of the tracks and require that they be larger than zero. While this strategy uses well understood collider techniques, it adds one more layer of complexity to relate the impact parameter to the displacement distance.

An alternative possibility is to use the variable called $L_{xy}$ defined in~\cite{CMS:2014wda}. If there is only one long-lived particle decaying in a region of the detector, then all the tracks that come from that decay will intersect at one point, and this point is the reconstructed displaced vertex. The distance away from the origin of this point in the $x-y$ plane is then $L_{xy}$.  In~\cite{CMS:2014wda} it was demonstrated that this method of reconstruction works well for two well separated long-lived particles at CMS. Extending this method to the case of many vertices in a relatively small space within the detector may be more challenging, but the high density of different detector channels could make it possible. From now on, we will assume that it is possible to reconstruct the vertex of the tracks using either the impact parameter or the $L_{xy}$ method. This allows us to discriminate emerging jets from the more common ones.

\begin{figure}
\centering
\includegraphics[width=1.0\textwidth]{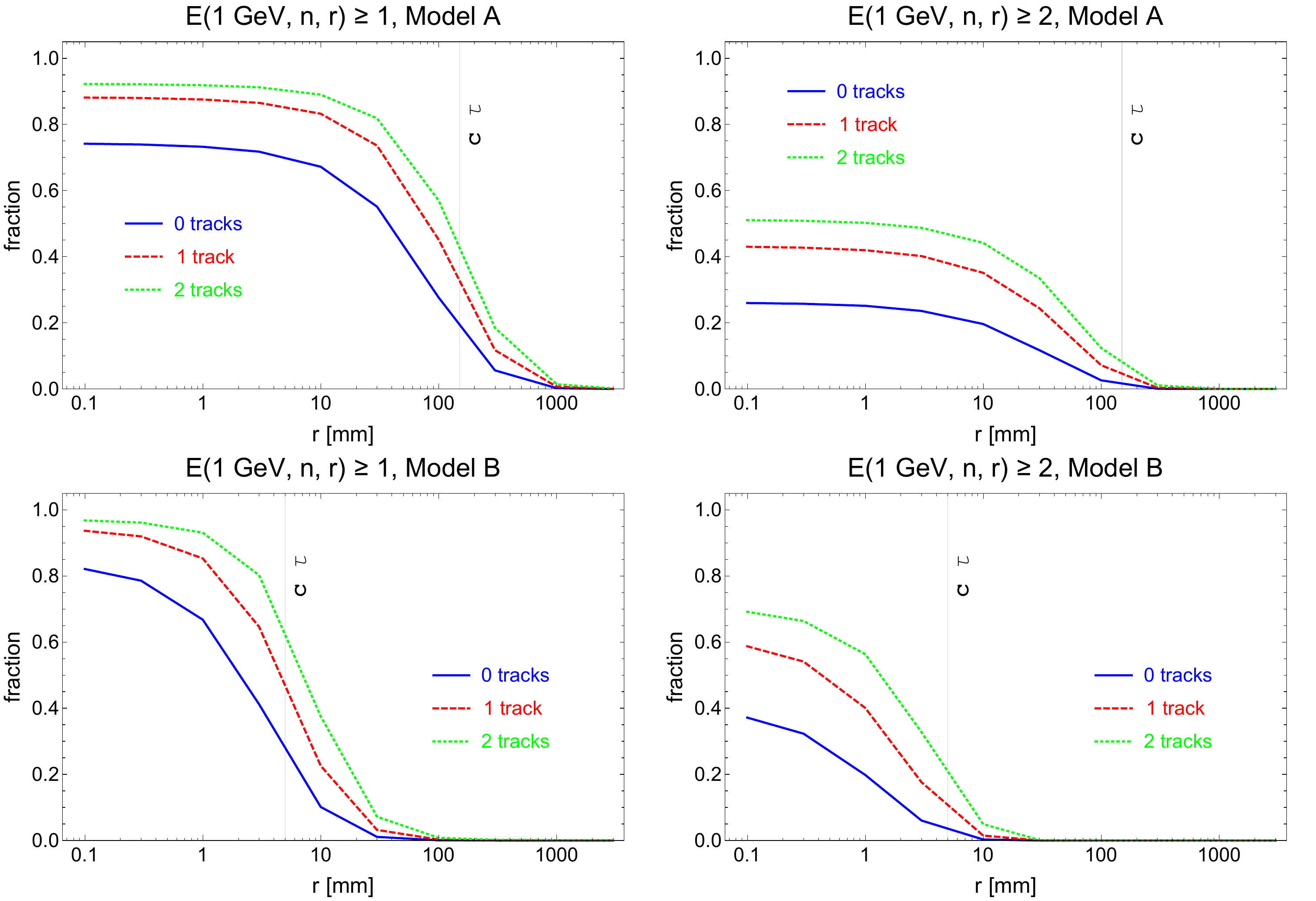}
\caption{Fraction of signal events in model A (top) and model B (bottom) which have at least one (left) or two (right) emerging jets with $p_T^{\rm min} = 1$ GeV as a function of $r$, the transverse distance. Within each plot, the curves are a maximum of 0, 1, and 2 tracks with transverse origin less than $r$ going from bottom to top. The vertical lines indicate the dark pion proper lifetimes $c\tau_0 = 150$~mm (5~mm) for model A (B).  
%A vertical line and the label $c\; \tau$ is placed at the proper lifetime of the dark pions for each particular model. 
All events must pass the kinematic preselection cuts. }
\label{fig:trackless-sig}
\end{figure}

We can now analyze the signal using our new emerging jet reconstruction object. In Fig.~\ref{fig:trackless-sig} we plot the fraction of signal events that contain at least one or two emerging jets for the two different benchmarks. Inspired by~\cite{CMS:2014wda}, we have taken $p_T^{\rm min} = 1$ GeV to avoid soft tracks. We see that for $r$ much less than the lifetime, nearly all events have at least one emerging jet and about half have two or more. We also see that the efficiency only moderately decreases with decreasing number of tracks $n$. We have not simulated pile-up here which could affect the results, and we will discuss possible mitigation strategies below. 

\begin{figure}
\centering
\includegraphics[width=.6\textwidth]{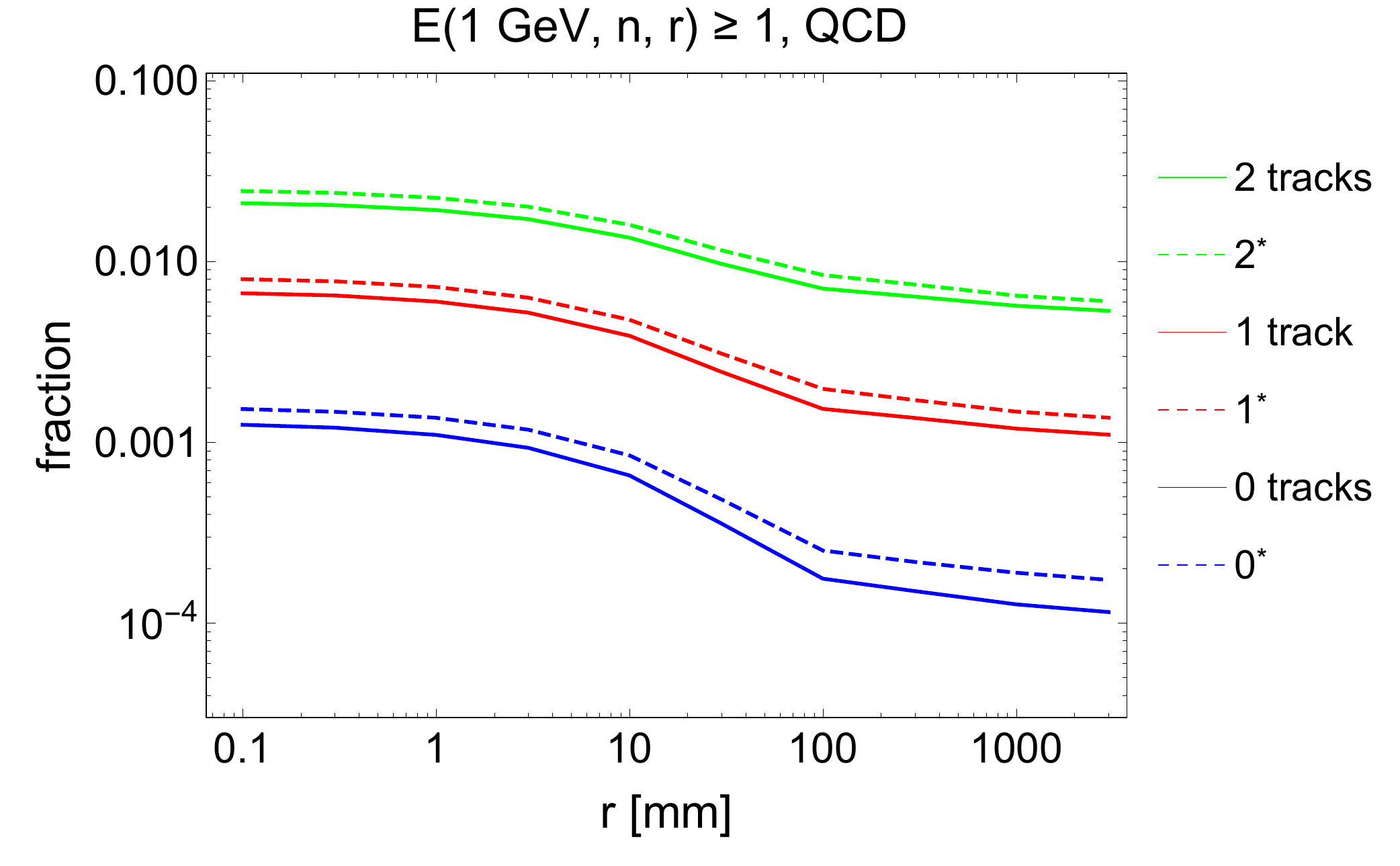}
\caption{Fraction of 4-jet QCD events that have at least one emerging jet as a function of the radius, $r$. These events have the kinematic cuts already applied, see text. From bottom to top, the lines are emerging jets with at most 0, 1, and 2 tracks inside of the radius $r$. The solid lines use the standard \textsc{Pythia} tune, while the dashed lines are the modified tune designed to increase the number of emerging jets in the sample~\cite{peter}. }
\label{fig:trackless-bg}
\end{figure}

Next we make a plot analogous to Fig.~\ref{fig:trackless-sig} for the QCD background. This is shown in Fig.~\ref{fig:trackless-bg} for events which have at least four jets and pass the kinematic cuts. We see that even allowing two prompt tracks in the jet eliminates more than 95\% of events, and requiring fewer tracks can do even better. We also see that it is relatively insensitive to the radius chosen, but that there is a drop-off around 50 mm where the majority of $b$ hadrons have decayed. 

Our background analysis  depends on the number and type of hadrons produced in the QCD events. This is not calculated from first principles in QCD, and is instead modeled in Monte Carlo programs such as \textsc{Pythia}. To get a sense of the size of this uncertainty arising from this, we compare the output of \textsc{Pythia} with the standard tune to a modified tune described in detail in Appendix~\ref{app:modifyPythia}~\cite{peter}. This tune is designed to enhance the number of jets with a small number of hadrons which makes it easier to have jets with very few charged tracks. The tune also enhances strangeness of the jets in order to have more hadrons with long lifetimes. The fraction of events which pass the kinematic cuts for the two different tunes are nearly identical, giving us confidence that changing the tune does not modify the gross kinematic structure of the events. We have also checked that the distributions in Fig.~\ref{fig:bgComp} are quite similar for the modified tune. The fraction of events with emerging jets in the modified tune are shown with dashed lines in Fig.~\ref{fig:trackless-bg}, and we see that while the fraction of trackless jets is increased, the effect is small. 

Putting all the elements together we show an example cut flow in Tab.~\ref{tab:cut-flow4}. We see that having just one emerging jet dramatically improves the signal to background ratio, but having two can bring this to a nearly background free search. In the twenty million background events we generated, there were only four events with two emerging jets for $r=3$ mm, and zero events with more than one emerging jet for $r=100$~mm. We can therefore estimate an upper bound on the background cross section and find it to be very small.

\begin{table}[tb]
\centering
\begin{tabular}{|c|c|c||c|c|}
%Signal numbers updated Feb 3
%BG numbers updated Jan 14
\hline
 & Model \textbf{A} & Model \textbf{B}  & QCD 4-jet & Modified \textsc{Pythia}\\ \hline 
Tree level & 14.6  &  14.6 & 410,000 & 410,000 \\ \hline
$\geq$ 4 jets, $|\eta| < 2.5$ & & & &  \\
$p_{T}$(jet) $> 200$ GeV & 4.9  &  8.5 & 48,000 & 48,000 \\
$H_T > 1000$ GeV &  & & & \\ \hline\hline
%$\Delta m/\bar{m} < 0.1$ &   &  & \\ \hline
$E(1 {\rm \,GeV}, 0,  3 \,{\rm mm}) \geq 1$ & 3.6  & 3.5 & 45 & 57 \\ \hline
$E(1 {\rm \,GeV}, 0,  3 \,{\rm mm}) \geq 2$ &  1.2 & 0.5 & $\sim 0.08$ & $\sim 0.04$ \\ \hline\hline
$E(1 {\rm \,GeV}, 0,  100 \,{\rm mm}) \geq 1$ & 1.4  &  $\lesssim 0.01$ & 8.5 & 12 \\ \hline
$E(1 {\rm \,GeV}, 0,  100 \,{\rm mm}) \geq 2$ &  0.1 & $\lesssim 0.01$ & $ \lesssim 0.02$ & $ \lesssim 0.02$ \\ \hline
\end{tabular}
\caption{Cut flow of the four jet analysis. Numbers in columns are cross sections in fb at LHC14. For the signal we take the mass of the bifundamental $M_X = 1$ TeV.  The two right most columns are different background estimates, the first using the standard \textsc{Pythia} tune, while the second uses the modified tune~\cite{peter}. The tree level cross section for the background is with the generator level cuts discussed in the text. }%%
\label{tab:cut-flow4}
\end{table}

\

\begin{figure}
\centering
\includegraphics[width=.45\textwidth]{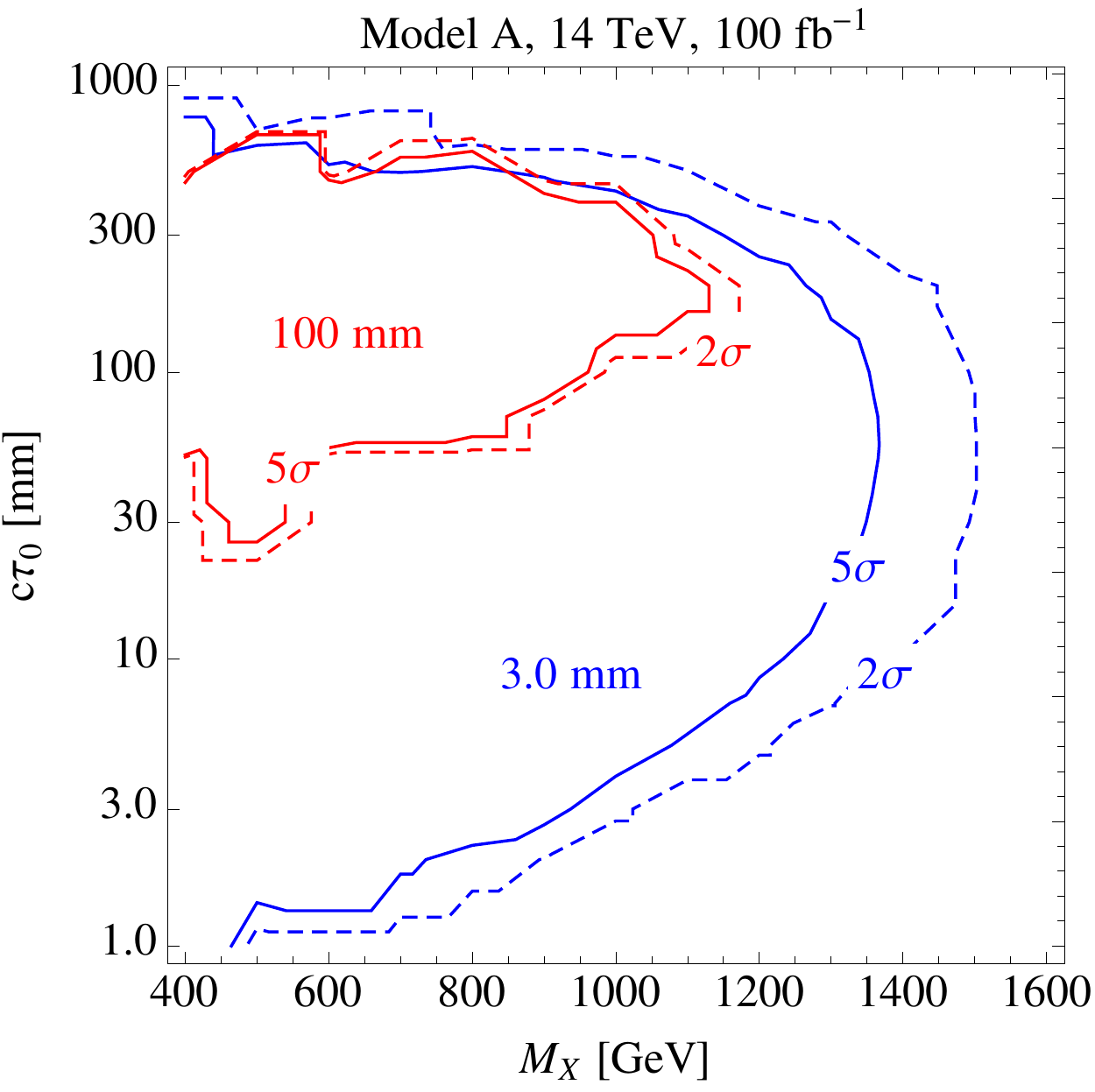}
\hspace*{0.5cm}
\includegraphics[width=.45\textwidth]{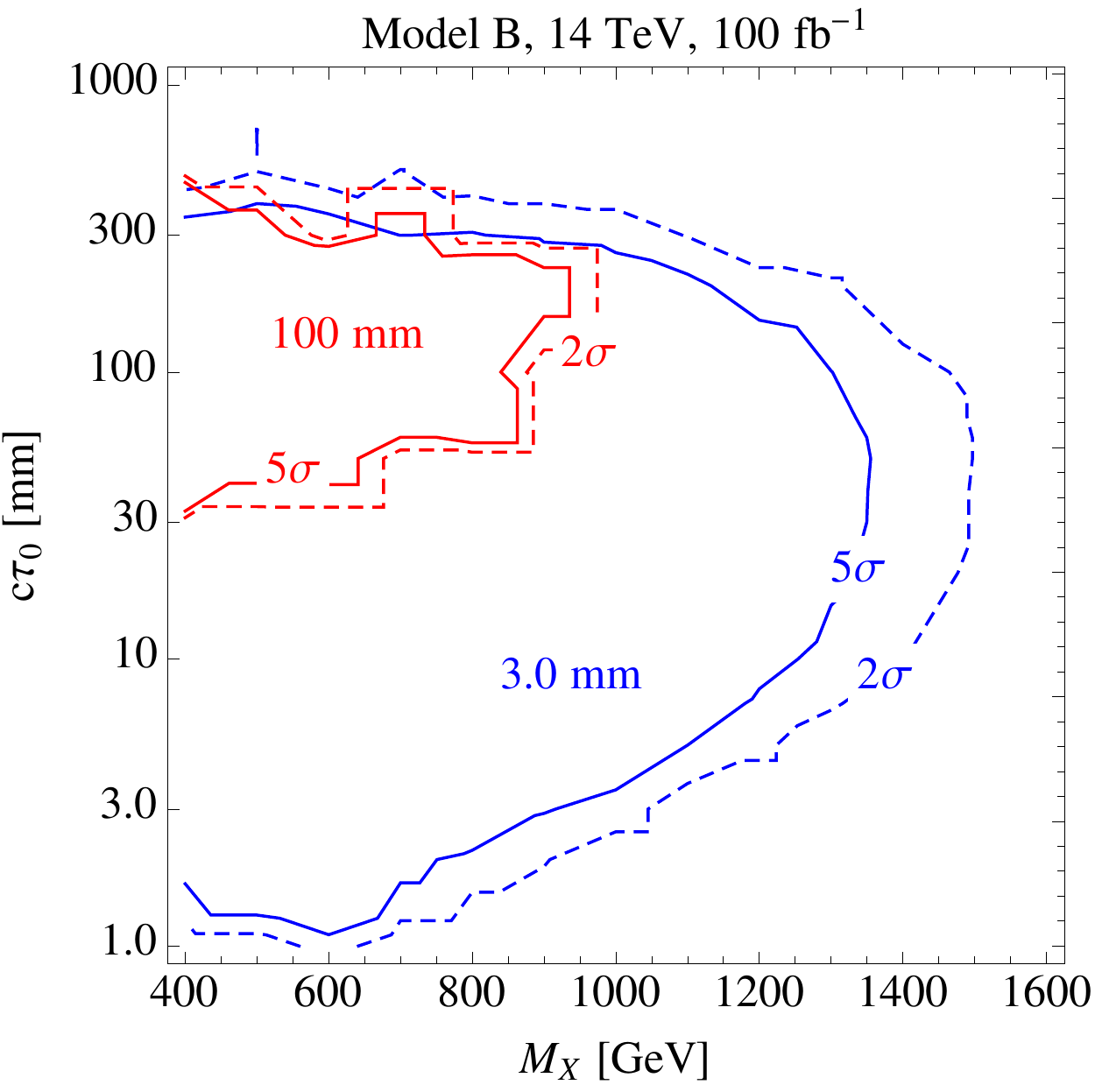}
\\[.2cm]
\includegraphics[width=.45\textwidth]{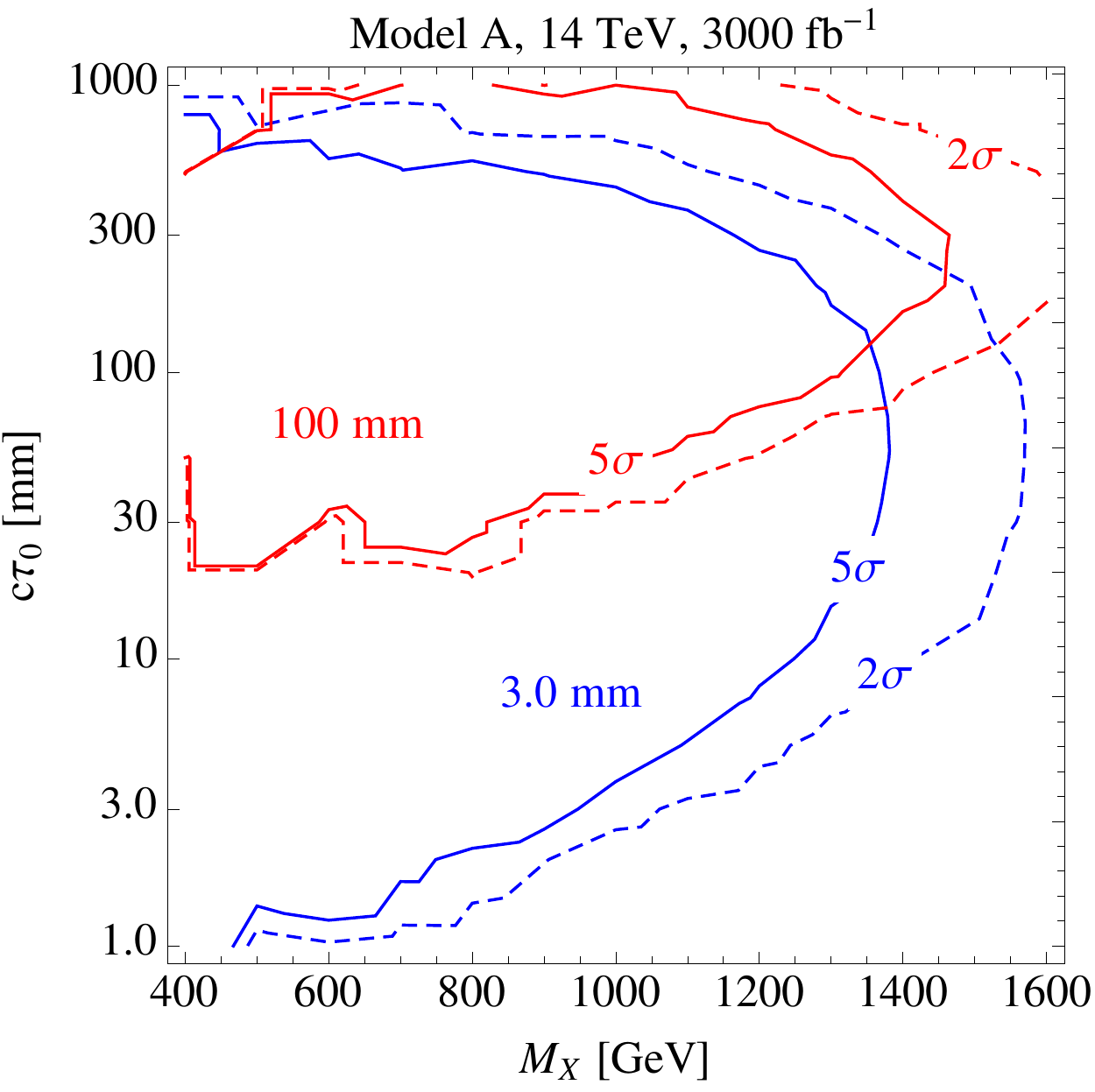}
\hspace*{0.5cm}
\includegraphics[width=.45\textwidth]{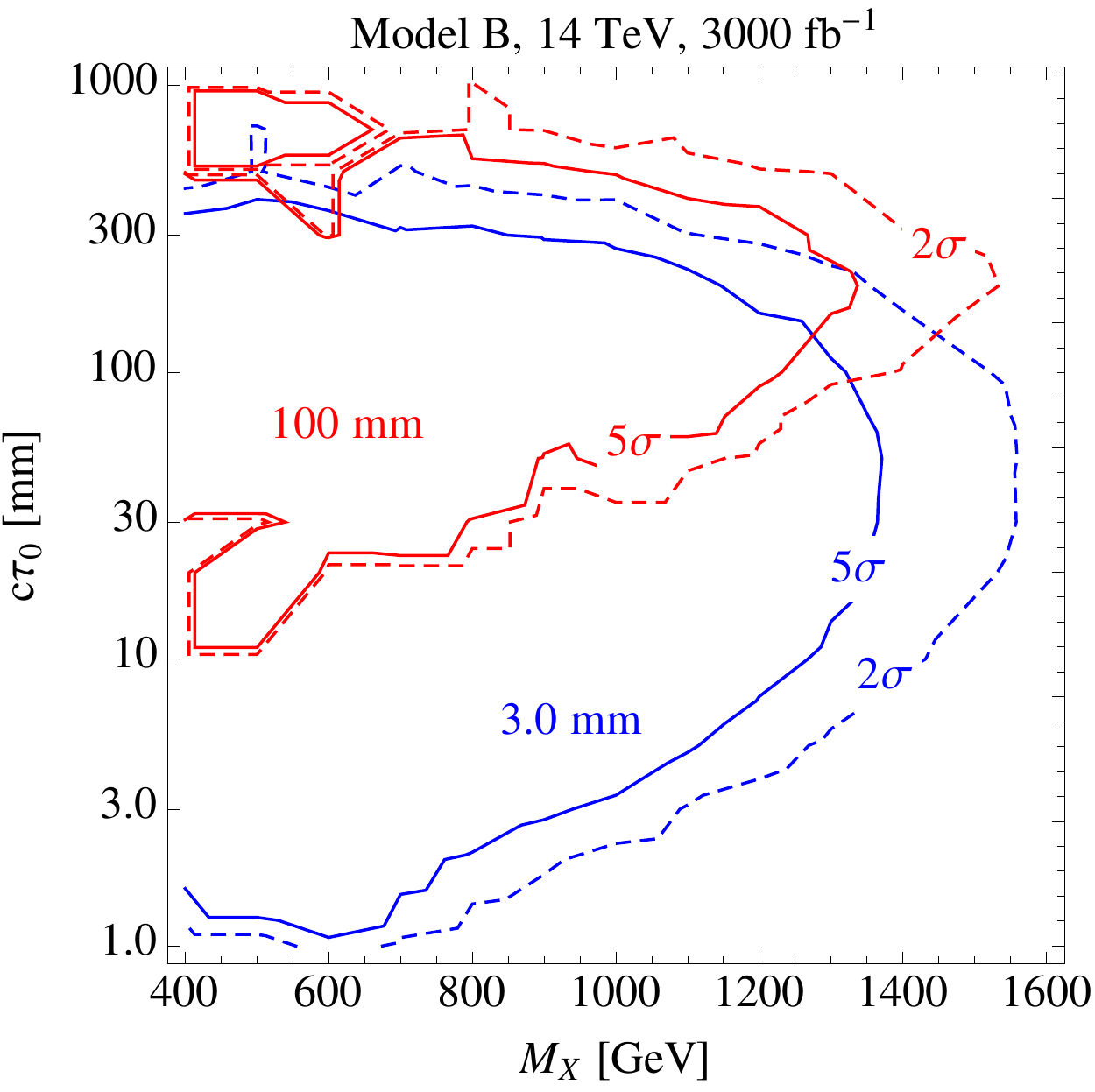}
\caption{Region of lifetime and mediator mass parameter space probed with 100~fb$^{-1}$ (top row) and 3000~fb$^{-1}$ (bottom row) at the 14~TeV LHC. For each model we show $2\sigma$ (dashed) and $5\sigma$ contours (solid) in the $M_X - c\tau_0$ plane, assuming a systematic uncertainty of 100\% on the background. The different colors correspond to requiring $E(1 {\rm \,GeV}, 0,  3 \,{\rm mm}) \geq 2$ (blue) and $E(1 {\rm \,GeV}, 0,  100 \,{\rm mm}) \geq 2$ (red). 
}
\label{fig:reachAB}
\end{figure}

We can now obtain the reach of the 14~TeV LHC. The significance is estimated using %
\begin{align}
	\sigma = \frac{S}{\delta B} \approx \frac{S}{\sqrt{B + \beta^2 B^2} } \,,
\label{eqn:significance}
\end{align}
where $\beta$ is the systematic error on the background estimate, and we use $\beta=100\%$ in the following. In addition we require $S>10$, otherwise we set $\sigma=0$.  The largest sensitivity always comes from the signal regions with two emerging jets, so we only present the reach in those channels. In Fig.~\ref{fig:reachAB} we show the region of parameter space that can be probed with 100~fb$^{-1}$ and 3000~fb$^{-1}$ at 14~TeV. For both models we vary the mediator mass $M_X$ and the proper lifetime of the dark pions, $c\tau_0$. 
The $r=3$~mm cut performs better in most regions of parameter space, and more than two orders of magnitude in lifetime can be probed, with exclusion being possible for mediator masses up to $1.5$~TeV. Sensitivity is lost when either the lifetime becomes too short, so that no signal events pass the emerging jet cuts, or when the lifetime becomes too large, in which case most dark pions decay outside of the calorimeter, and the events fail the kinematic cuts. Both cases could be improved by putting the emerging jet cuts even closer to the interaction point and by including dark pions which decay in the muon system in the jet reconstruction. 

Models A and B differ mainly in the mass spectrum. The lighter states of model B are more boosted on average and therefore are more likely to decay outside of the calorimeter given the same $c\tau_0$, which explains the lower sensitivity in the large $c\tau_0$ region compared to model A. Furthermore the larger multiplicity of dark pions in model B makes it more likely for some of them to decay early, therefore causing events to fail the emerging jet cuts. The $p_T$ weighted strategy which we outline in the next section could lead to improvements here and for models with even lower dark pion masses. Instead larger dark pion masses should not have an adverse effect on the sensitivity, at least until we reach a point where most of the jet energy is contained in a single massive dark pion, in which case displaced dijet searches could be more sensitive. 

The 100~mm search is essentially background free, and thus the reach is limited by  production rate times acceptance. It follows that going from 100~fb$^{-1}$ to 3000~fb$^{-1}$ can significantly improve the reach in this case.\footnote{The background estimate for the $E(1 {\rm \,GeV}, 0,  100 \,{\rm mm}) \geq 2$ cut is limited by Monte Carlo statistics. To obtain a better estimate for the background in this channel, we use the square of the background suppression of the $E(1 {\rm \,GeV}, 0,  100 \,{\rm mm}) \geq 1$ cut, which gives an estimated background of 0.0015~fb (0.003~fb) using the default (modified) background simulation. While for a luminosity of 100~fb$^{-1}$ this doesn't affect the reach, it is relevant for the 3000~fb$^{-1}$ projection, where we use 0.003~fb for the background estimate. } Instead the 3~mm analysis is already limited by $S/B$ and does not benefit so much from the increased luminosity. On the other hand it is certainly possible to optimize this search for the high-luminosity run by rejecting backgrounds more aggressively, and by reducing the uncertainty on the background. One could also imagine asking for a third emerging jet which can originate from a hard splitting in the dark sector. 

Pileup could potentially reduce the signal efficiency of our analysis. The well reconstructed primary vertex of the two QCD jets should allow efficient discrimination of pileup events, such that their tracks will not be counted. We therefore did not include pileup in our simulation. Multi-parton interactions instead will produce tracks originating from the same vertex, and have been included in the simulations for signal and backgrounds. A strategy to further reduce possible effects of pileup is discussed in the next section.

\subsection{Alternative Strategy: $p_T$ Weighting}

In this section we present an alternative based on using the $p_T$ fraction of the jet which is emerging, rather than counting tracks. As before, this requires reconstruction of displaced charged tracks in order to determine $L_{xy}$, how far from the origin in the $x-y$ plane they originate. This strategy, however, is more robust to pileup because while a pile up event can produce tracks above the 1 GeV threshold from the previous section, they are much more unlikely to make a substantial contribution to the $p_T$ of a jet. 

For this section we define the displaced $p_T$ fraction $F(r)$ for a jet as a function of radius $r$ as:
\begin{equation}
F(r) = \frac{1}{p_T^{\rm calo-jet}} \sum_{L_{xy} > r} p_T^i
\end{equation}
where $p_T^i$ is the $p_T$ of charged tracks associated with the jet with $L_{xy} > r$ which we normalize to the calorimeter $p_T$ of the jet. This variable goes from 0 to 1 for a given jet. For QCD jets it tends to take values near zero since most of the energy is in prompt tracks. A jet can only have $F=1$ if it is composed entirely of charged tracks which originate further away than $r$. This is because neutral particles contribute to the denominator in the prefactor but do not contribute to the sum. By isospin conservation, we expect approximately half of the decay products of the dark mesons to be neutral, so we expect the $F$ distribution for signal jets to be peaked around 0.5 for $r$ less than the lifetime of the dark pions. 

\begin{figure}
\centering
\includegraphics[width=0.9\textwidth]{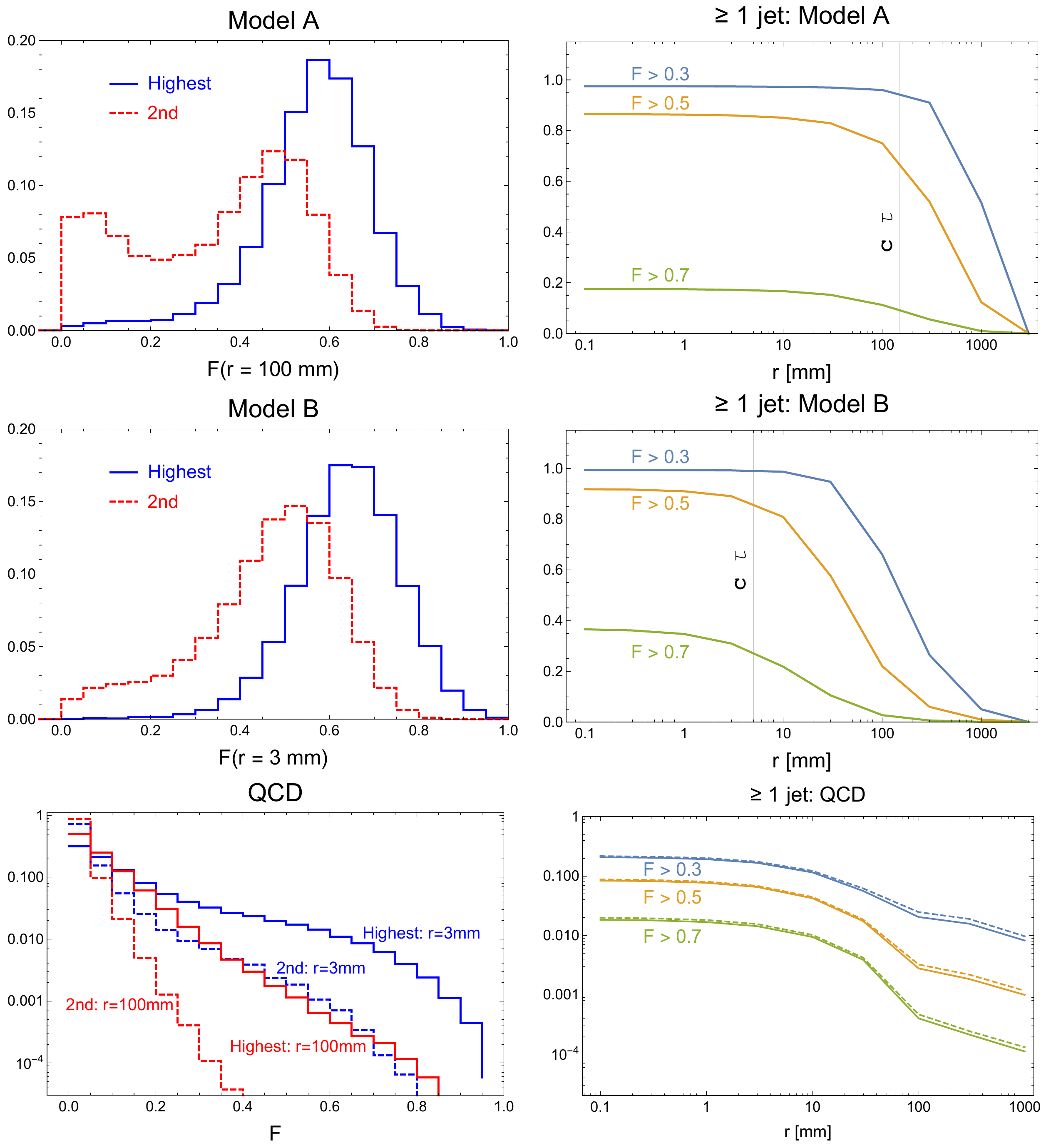}
\caption{$F$ distributions for model A (top), model B (middle), and QCD background (bottom). The left plots are the distribution of the highest and second highest $F$ values for jets in an event, where for model A (B) we have taken $r=100$ (3) mm, and for the background we show both. The right plot shows the fraction of events that have at least one jet with $F > $ 0.3, 0.5, or 0.7. All events must pass the kinematic cuts in Tab.~\ref{tab:cut-flow4}. Note that the signal plots use a linear scale while the background plots use a log scale, and the dashed lines in the bottom right plot are those using the modified \textsc{Pythia} tune.  }
\label{fig:ptSig}
\end{figure}

We now analyze this variable more quantitatively, with the main results of this section given in Fig.~\ref{fig:ptSig}. The top two rows show distributions for signal, and we see that for emerging jets the distributions do peak around 0.5 with very few jets having $F$ near one. For model A a non-negligible fraction of events have only one emerging jet. This comes from one of the signal jets being too soft or too forward, and the extra jet to pass the kinematic cuts coming from splitting and/or ISR. From the plots in the right column we see that the fraction of events that will pass any cut is insensitive to $r$ for $r$ smaller or comparable to the lifetime. For larger $r$, the efficiency decreases slowly because the highest energy pions tend to be the ones that travel the further because of relativistic boost. Therefore, even for distances much larger than the proper lifetime, there is still a reasonable fraction of events that pass this cut. This contrasts with the emerging jet definition depicted in Fig.~\ref{fig:trackless-sig}, where there is a much steeper drop as a function of $r$ because we only require one dark pion to decay at a radius less than $r$.

We now turn to the QCD background quantified in the bottom row of Fig.~\ref{fig:ptSig}. We see that the $F$ distribution is peaked at zero and steeply falling. We also see that it is much more steeply falling for $r=100$ mm than for 3 mm. This is a consequence of $b$ hadrons; in Fig.~\ref{fig:bgComp} we see that $b$ hadrons tend to decay between 1 and 100 mm, so for $r=3$ mm, there will be many undecayed neutral $b$ mesons that will contribute to $F$, but for  $r=100$ mm, only strange mesons contribute. Looking at the bottom right plot we see that there is a strong break, and going to $r=100$ can give QCD rejection $\mathcal{O}(10^3)$ by requiring one jet with large $F$, and much better if we require two such jets. 

When we analyze the signal and background together, we find that using $r=3$ mm there is a very large background from $b$ hadrons so it is impossible to sufficiently reduce the background without killing the signal. The experiments, however, are very good at finding $b$ jets, so using those techniques it is likely possible to distinguish the $b$ background from the signal using not only lifetime information but also invariant masses and decay products. Because of the complexity of experimental $b$-tagging algorithms, we cannot simulate them here, but we stress large improvements may be possible. 

Instead we will focus on $r=100$ mm where the $b$'s have mostly decayed and the strange background is much smaller. This method works for model A with the long lifetime, but there is even marginal sensitivity to model B with a much shorter lifetime. We show an abbreviated cut flow in Tab.~\ref{tab:pt-cut-flow} for mediator mass of 1 TeV, and we see that requiring two jets with $F>0.5$ leads to a signal to background ratio much larger than one, allowing a possible discover at the LHC.

\begin{table}[tb]
\centering
\begin{tabular}{|c|c|c||c|c|}
\hline
 & Model \textbf{A} & Model \textbf{B}  & QCD 4-jet & Modified \textsc{Pythia}\\ \hline 
$\geq$ 4 jets, $|\eta| < 2.5$ & & & &  \\
$p_{T}$(jet) $> 200$ GeV & 4.9  &  8.5 & 48,000 & 48,000 \\
$H_T > 1000$ GeV &  & & & \\ \hline\hline
1 jet $F(100$ mm) $> 0.5$ &  3.7 & 1.9 & 130 & 150\\ \hline
2 jets $F(100$ mm) $> 0.5$ & 1.2 & 0.1 & 0.2 &  0.2\\ \hline\hline
$\sigma(100\, {\rm fb}^{-1})$ & 5.9 & 0.5 & - & - \\ \hline
\end{tabular}
\caption{Same as Tab.~\ref{tab:cut-flow4} but with $p_T$ weighted variables. The last row shows the discovery significance $\sigma$ defined in Eq.~(\ref{eqn:significance}) again taking $\beta=100\%$. }%%
\label{tab:pt-cut-flow}
\end{table}

We present this alternative method, because unlike the one in Sec.~\ref{sec:emergingjets}, it is an affirmative search for the emerging property. The previous method uses the fact that prompt tracks are a feature of the background and requires the absence of them. This allows backgrounds such as jets of neutrons and/or photons, which are not signal-like at all. The current method is an affirmative search for the emerging property, namely a search for energy which emerges at large transverse distances. Therefore the background must look much more like the signal to pass the cuts. The other advantage of this method is that it is much more insensitive to detector effects such as cosmic rays and pileup. Pileup in particular, can add one track to a jet which would be enough to make it not emerging. On the other hand, pileup cannot make an $\mathcal{O}(1)$ change in the energy dynamics of a jet, thus making this method very robust to the high pile up environment of the high luminosity LHC.

\section{Prospects at LHCb}
\label{sec:lhcb}
\begin{figure}
\centering
\includegraphics[width=0.55\textwidth]{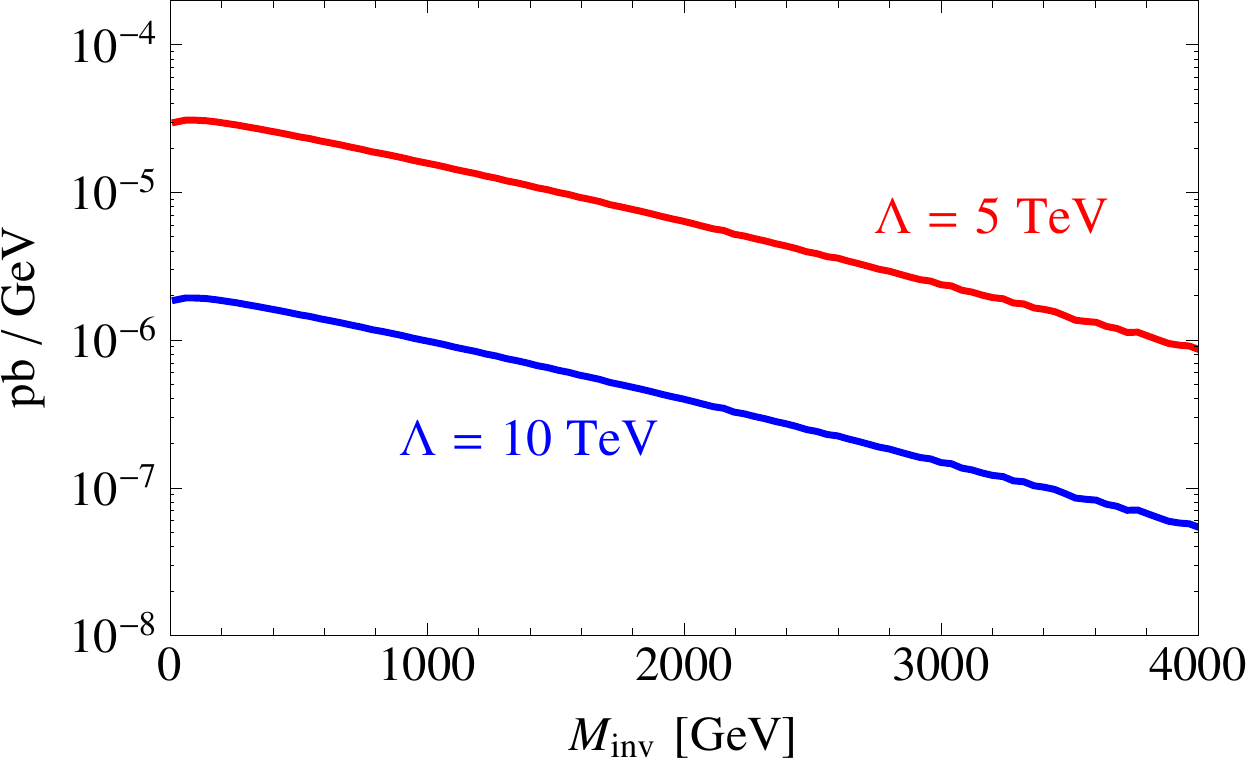}
\caption{Dark quark invariant mass distribution for different values of the cut-off $\Lambda$ at the 14~TeV LHC. The total integrated cross section for the process $pp \to \bar{Q}_d Q_d$ is $42$~fb for $\Lambda=5$~TeV and $2.5$~fb for $\Lambda=10$~TeV per dark quark flavor, for $N_d = 3$.  } 
\label{fig:minv}
\end{figure}

Our proposed analyses for the ATLAS and CMS detectors rely on on-shell production of heavy mediators, whose decay give rise to emerging jets. The reach of those searches is limited by the kinematic reach of the LHC experiment. However even if the mediators are too heavy to be produced directly at the LHC, dark quark pairs can still be produced through effective operators of the form
\begin{align}
	{\cal L}\supset  \frac{1}{\Lambda^2}(\bar{q} \Gamma_q q ) (\bar{Q}_d \Gamma_d Q_d )\,,
\end{align}
with appropriate Dirac structures $\Gamma$. We already made use of such an operator in Sec.~\ref{sec:darkpions} to understand the decays of dark pions. As we can see from Fig.~\ref{fig:minv}, the differential cross section peaks at very low invariant mass, so events induced by these operators tend to have small $H_T$ and would be difficult to trigger on at ATLAS and CMS. Nevertheless they can lead to sizable production rates for dark pions. The idea would then be to search directly for these dark pions in the LHCb detector from their decay to SM mesons. 

Reconstructed dark pions can be differentiated from SM mesons by their invariant mass, by their lifetime and by their decay products and branching ratios. While a full simulation is beyond the scope of this paper, in the following we will estimate the event rate that can be expected at LHCb and show some kinematic properties of the produced dark pions.
For definiteness, we will consider the operator ${\cal O}_u = 1/\Lambda^2 (\bar{u} \gamma_\mu u ) (\bar{Q}_d \gamma^\mu Q_d )$, which can originate from integrating out either a $Z'$ boson or a bi-fundamental scalar, as discussed in Sec.~\ref{sec:models}. Coupling to $\bar u u$ yields the largest cross sections, which should give the strongest constraints. At the 14~TeV LHC, we find
\begin{align}
	\sigma (p p \to \bar{Q}_d Q_d ) \approx (8.2~{\rm pb}) \times N_d \times n_f \times \left( \frac{\rm TeV}{\Lambda} \right)^4
\end{align}
for the tree level cross section (with a cut of $\sqrt{\hat{s}} > 50$~GeV), which scales as $1/\Lambda^4$, as long as the EFT description is valid. If instead we consider the operator from Eq.~(\ref{eqn:effOp}) with $\Lambda = \kappa/ M_{X_d}$, the cross section is about a factor 8 smaller due to the smaller down quark PDFs and due to the chiral structure of the couplings. 

When comparing with the direct on-shell production of mediators, a few comments are in order. First, if we consider a $t$-channel mediator like $X_d$, the on and off-shell contributions are independent of each other, and controlled by different parameters. The direct production of the mediator is fully determined by the QCD coupling. The off-shell production of $Q_d$ pairs can be larger, but it is important to realize that it now has to compete with QCD dijet production, and it is unclear how an emerging dijet signal could be triggered on efficiently at ATLAS and CMS. 

If instead the operators would originate from integrating out a $Z'$ boson, the on-shell production and effective operator would contribute to the same final state, and direct $Z'$ production could easily dominate. Still as far as LHCb is concerned, the effective operator description is sufficient, since only part of the event is reconstructed, and we are mostly interested in the fraction of events where one or more dark pions enter the LHCb detector.\footnote{Additional care would be necessary in order to convert a limit on $\Lambda$ into a bound on the $Z'$ mass, since that limit will depend on the couplings and branching ratios of the $Z'$ as well as on the relative contributions of on and off-shell production of $Q_d$, due to the scaling of the produced dark meson number with $\sqrt{\hat{s}}$. }

\begin{figure}[tb]
\centering
\includegraphics[width=0.465\textwidth]{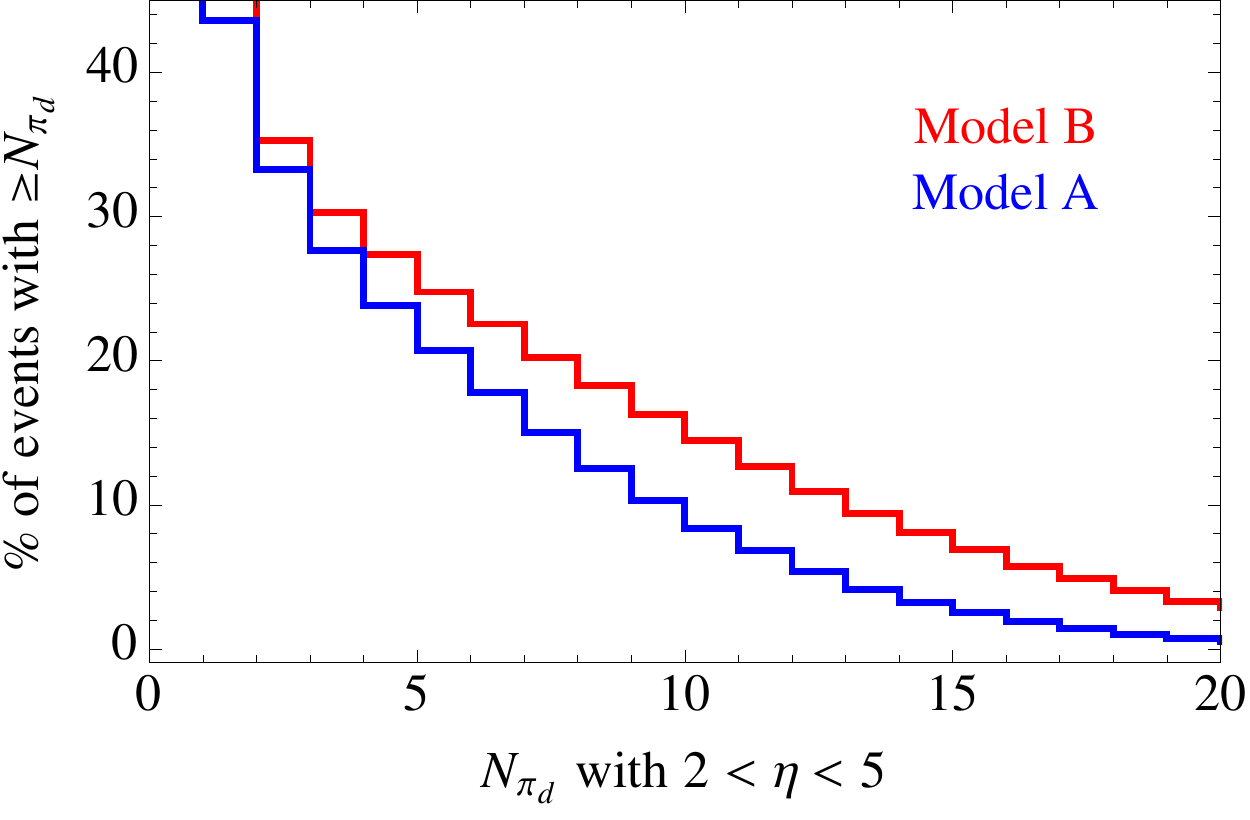}
\hspace*{.5cm}
\includegraphics[width=0.45\textwidth]{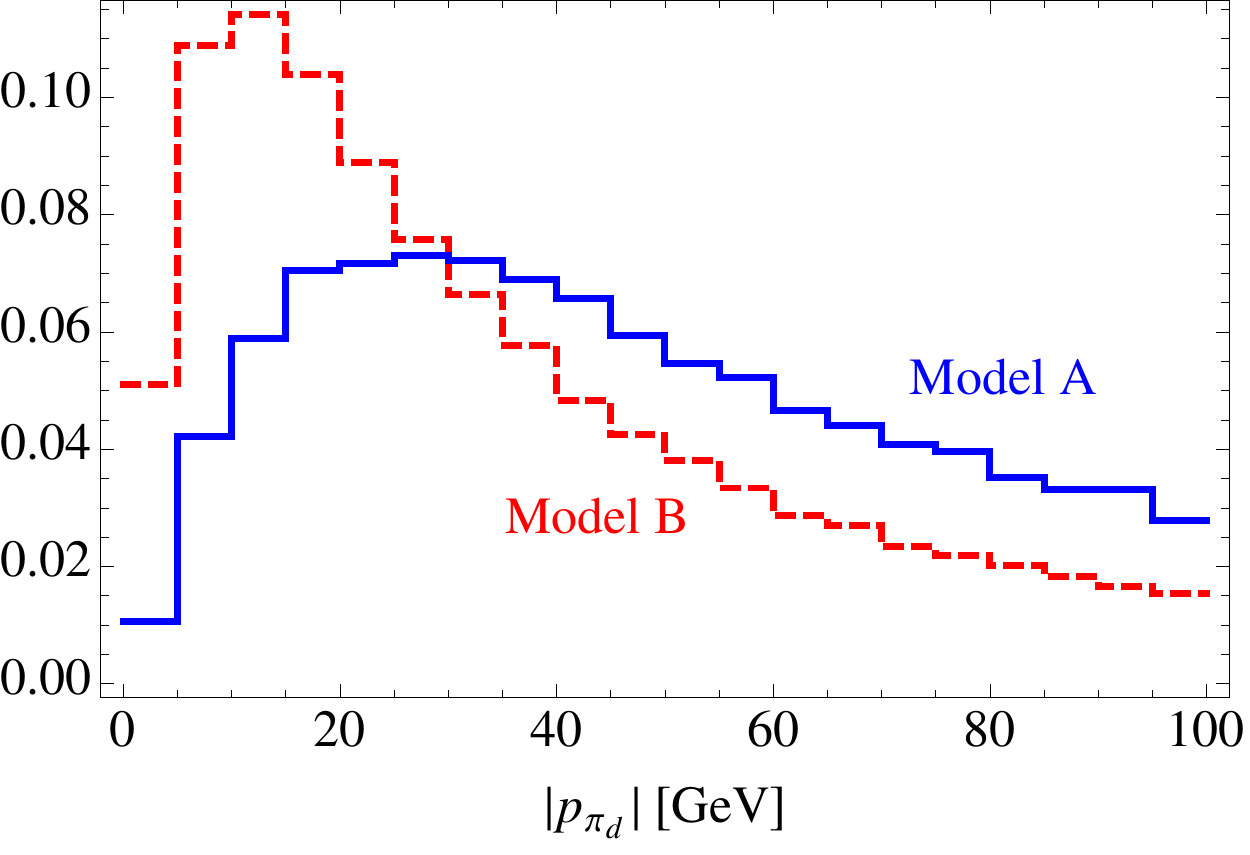}
\caption{Left: Fraction of $Q_d \bar{Q}_d$ events with at least $N_{\pi_d}$ dark pions inside the LHCb detector. About 45\% of all events have at least one dark pion in LHCb, and almost 30\% have three or more. Right: Momentum distribution of dark pions in the LHCb detector. } 
\label{fig:nLHCB}
\end{figure}

In Fig.~\ref{fig:nLHCB} we show the fraction of events where one or more dark pions end up in the LHCb detector. For both benchmark models, about half of all $Q_d \bar{Q}_d$ events have one or more dark pions in the pseudo-rapidity range of LHCb. Also shown is the momentum distribution of dark pions in the LHCb detector, where we see that model A produces a harder spectrum, due to the overall larger mass scale in that model.

\begin{figure}[tb]
\centering
\includegraphics[width=0.55\textwidth]{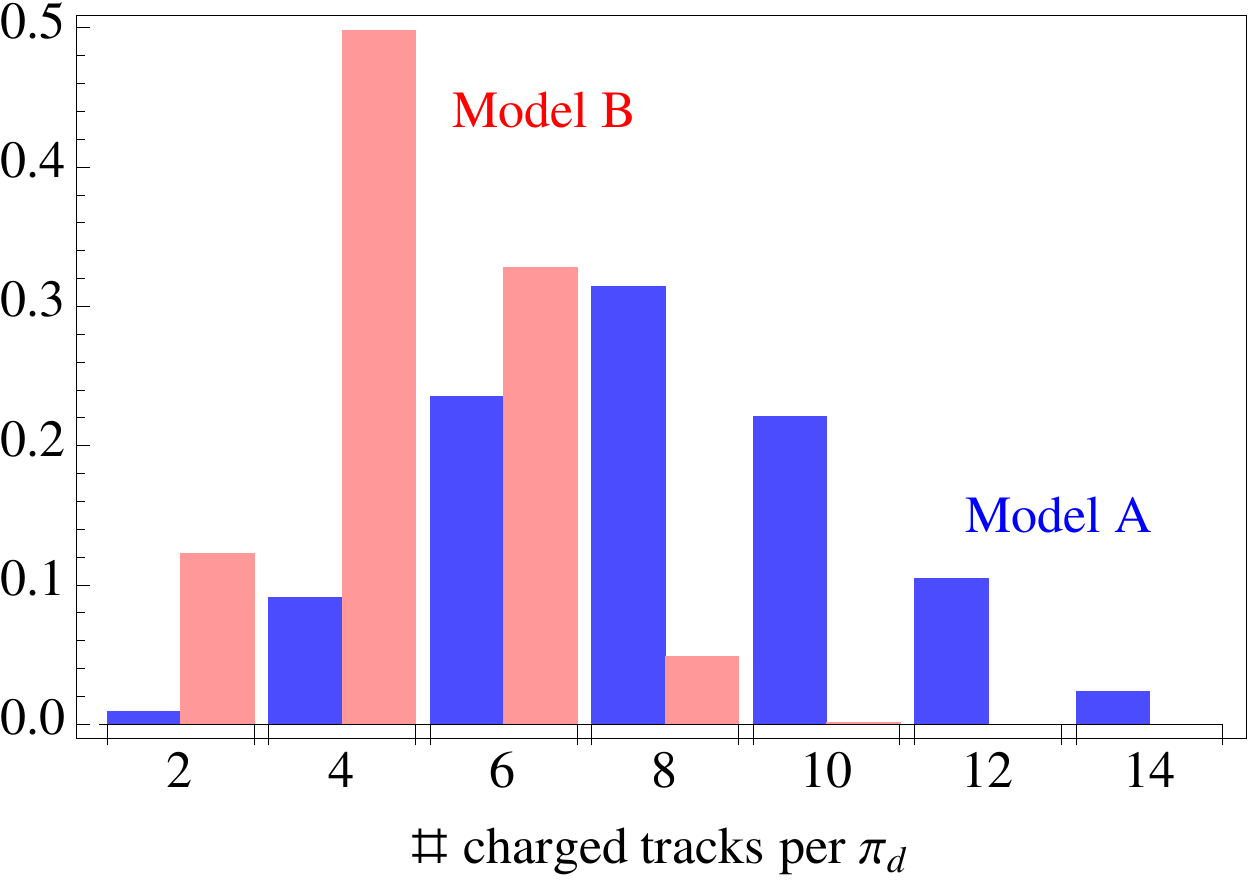}
\caption{Multiplicity of charged tracks in $\pi_d$ decays, assuming 100\% decay to down quarks, and with the fragmentation process simulated using \textsc{Pythia}. } 
\label{fig:decayModes}
\end{figure}

Obtaining precise predictions for the decay modes and branching ratios of $\pi_d$ to SM hadrons is difficult, since it depends on non-perturbative QCD fragmentation, as well as on the flavor structure of the couplings. In the \textsc{Pythia} implementation, those decays are simulated using the LUND string fragmentation model~\cite{Andersson:1998tv}, which is successful at modeling QCD fragmentation. For dark pion masses in the few GeV range, exclusive hadronic processes already become rare. Instead in order to get an idea about the characteristics of the signal, in Fig.~\ref{fig:decayModes} we show the multiplicity of prompt (with respect to the decay vertex) charged tracks from decays of dark pions. We see that up to 10 charged tracks appear regularly for the case of a $5$~GeV dark pion, while fewer tracks are expected for lighter $\pi_d$. For the figure we assume 100\% decays of dark pions into down quarks. If decays into heavier quarks would dominate, we would instead find fewer charged tracks, since for example charged kaons can carry away a larger fraction of the particle's rest mass. 

The trigger thresholds at LHCb~\cite{Albrecht:2013fba} are very loose when compared with ATLAS or CMS. At the level of the hardware trigger L0, a deposition of transverse energy $E_T$ of 3.7~GeV in the hadronic calorimeter or 3~GeV in the electromagnetic calorimeter are required. Next the high level triggers start with the reconstruction of tracks in the vertex locator (VELO).  In total a few tracks in the VELO and a moderate energy deposit in the calorimeters are enough for events to be recorded and analyzed.\footnote{We would like to thank Victor Coco for discussion on these points.} We can therefore expect that most events with one or more dark pions can be captured. Events with three or more reconstructed displaced dark pions might look sufficiently different from QCD backgrounds for the search to be background free. Then if we assume a reconstruction efficiency of 10\%, with 15~fb$^{-1}$ one could probe cross sections for $\sigma(p p \to {\bar{Q}_d Q_d}) $ as low as $10$~fb, corresponding to scales $\Lambda \sim 5$~TeV. While this is just a very crude estimate, the reach seems promising enough to warrant a more careful analysis.

\section{Sensitivity to Other New Physics Scenarios}
\label{sec:other}

Long lived particles decaying with displaced vertices are well motivated in many extensions of the SM. A well known example is the case of $R$-parity violating (RPV) supersymmetry~\cite{Barbier:2004ez}. Because the RPV couplings are in the superpotential, it is natural for them to be quite small, possibly small enough to make the LSP decay length macroscopic. 
Other more recent examples where displaced decays are motivated include displaced Higgs signatures~\cite{Chang:2005ht,Strassler:2006ri,Buckley:2014ika} or late Higgs production~\cite{Jaiswal:2013xra}, Lepton Jets~\cite{Falkowski:2010cm,Falkowski:2010gv} Baryogenesis~\cite{Barbier:2004ez,Cui:2014twa}, keV dark matter~\cite{Falkowski:2014sma}, heavy neutrinos~\cite{Helo:2013esa}, right-handed sneutrinos~\cite{Cerdeno:2013oya}, and twin Higgs models~\cite{Craig:2015pha}. 

When considering a specific model, a dedicated search will most likely deliver optimal results. For instance, if muons are likely to appear in the final state, those can be used for triggering purposes and to suppress backgrounds. On the other hand, given the variety of models on the market, it is also desirable to have searches which are more model independent, and thus will allow to place bounds on multiple new physics scenarios. 

In the following we will demonstrate that the emerging jet analysis can easily be used to obtain bounds on other new physics scenarios with displaced decays, even if their signature will appear different at first sight. As an example, we will use a supersymmetric scenario where the neutralino LSP decays through a $UDD$ type RPV operator.

\begin{figure}[tb]
\centering
\includegraphics[width=0.6\textwidth]{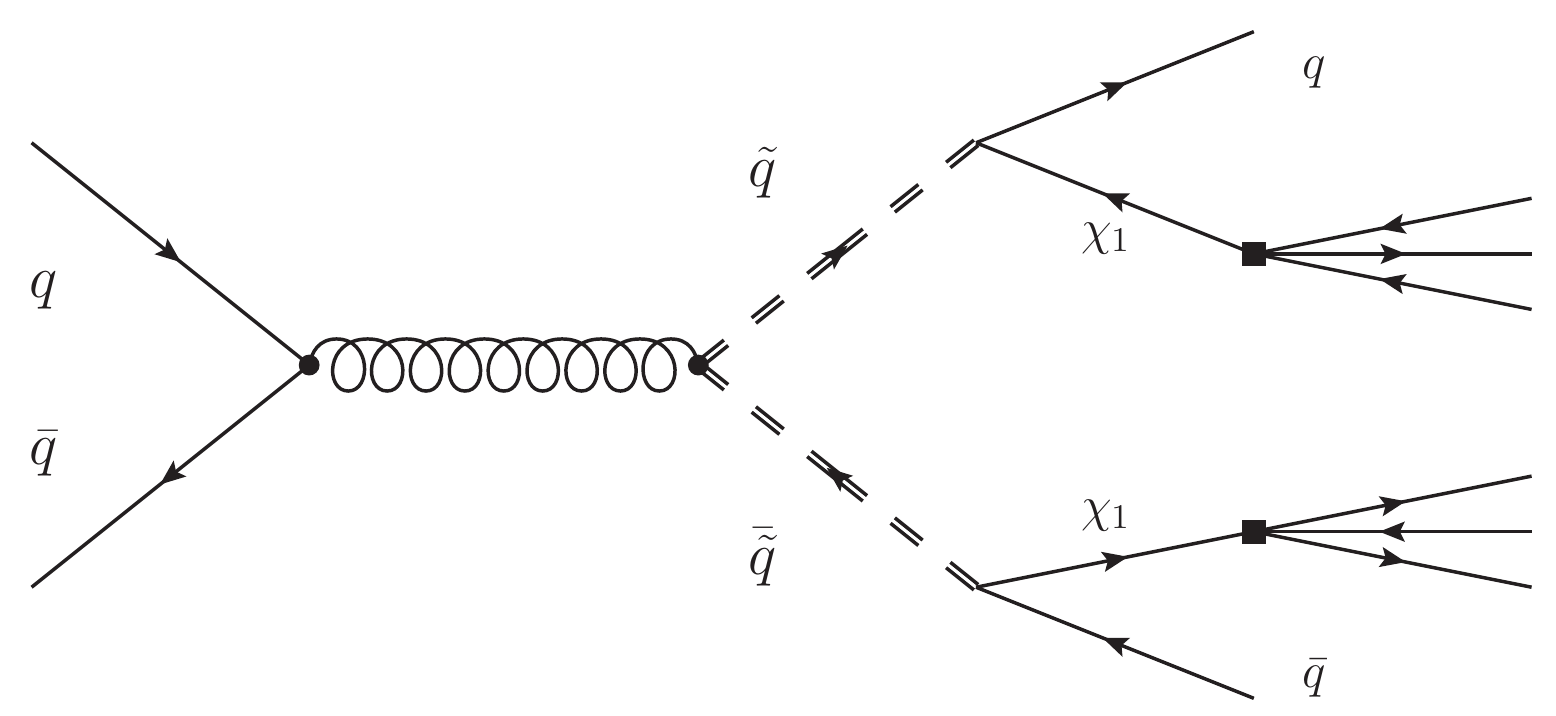}
\caption{Pair production of squarks $\tilde{q}$ with subsequent decay into quarks $q$ and neutralinos $\chi_1$. The neutralino undergoes an $R$-parity violating three-body decay into a $uds$ final state, and has a macroscopic lifetime. Not shown is the corresponding diagram with initial state gluons. } 
\label{fig:RPVneutralino}
\end{figure}

The process we have in mind is depicted in Fig.~\ref{fig:RPVneutralino}: squarks $\tilde{q}$ are pair produced and decay to a quark $q$ and the lightest neutralino $\chi_1$.  In the presence of $UDD$ type RPV operators, the lightest neutralino can undergo a three-body decay into three quarks, mediated by an off-shell squark. In the super potential, these operators can be written as~\cite{Barbier:2004ez}
\begin{align}
	{\cal W}_{\rm RPV} \supset \frac{1}{2} \lambda^{''}_{ijk} U_i D_j D_k\,,
\end{align}
where gauge invariance forces $\lambda^{''}_{ijk}$ to be anti-symmetric in $jk$. If the neutralino $\chi_1$ is the lightest supersymmetric particle (LSP), it can for example undergo the decay $\chi_1 \to uds$, mediated by an up or down-type squark. This decay is suppressed both by the squark masses and by the potentially small\footnote{See e.g.~\cite{Evans:2012bf,Bhattacherjee:2013gr} for currently allowed values of these couplings.} RPV couplings $\lambda^{''}_{ijk}$, and therefore $\chi_1$ may have a macroscopic decay length. The squarks, of course, decay promptly via gauge or Yukawa interactions: $\tilde{q} \to q \chi_1$.

\begin{figure}[tb]
\centering
\includegraphics[width=0.45\textwidth]{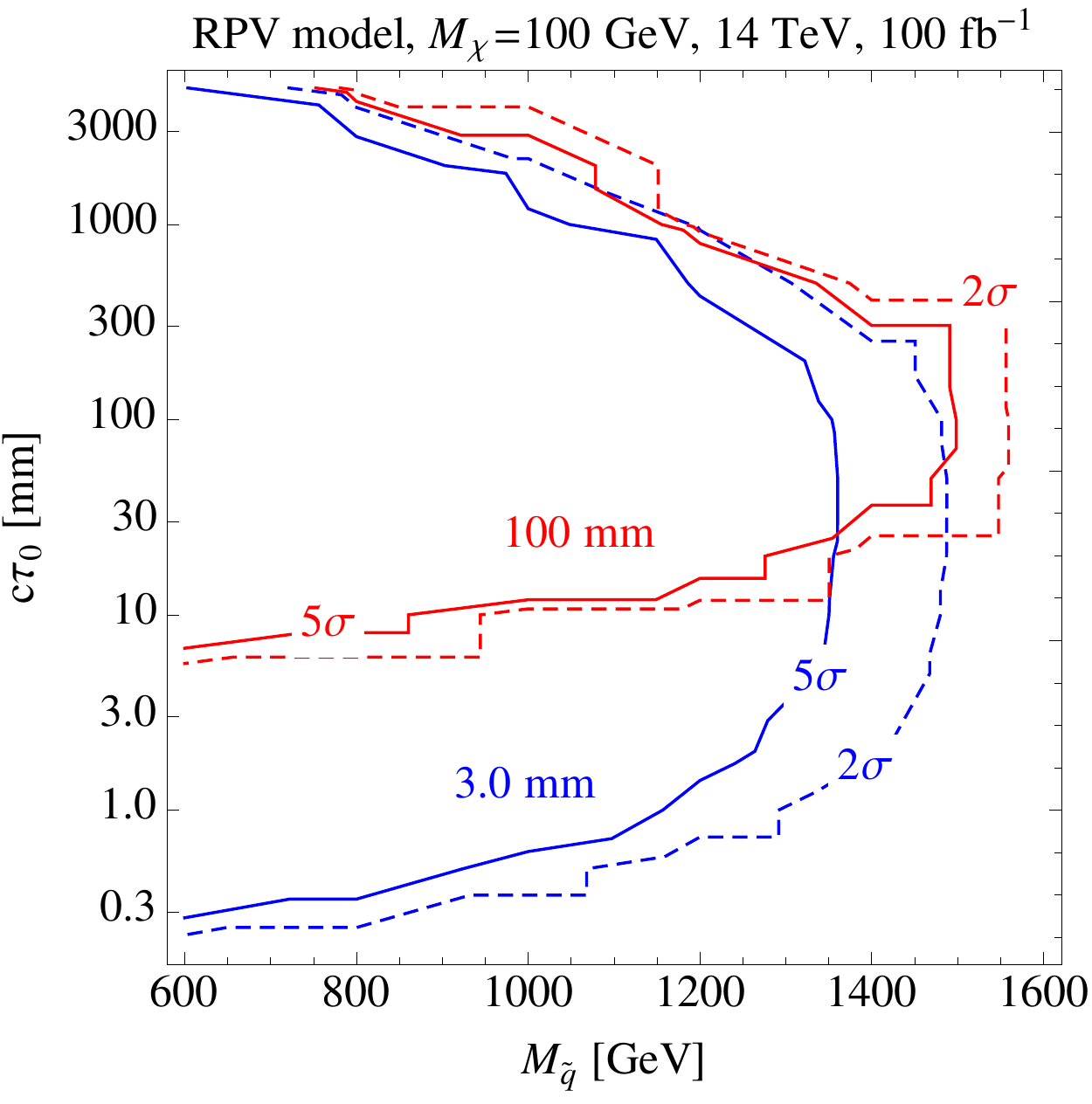}
\hspace*{0.5cm}
\includegraphics[width=0.45\textwidth]{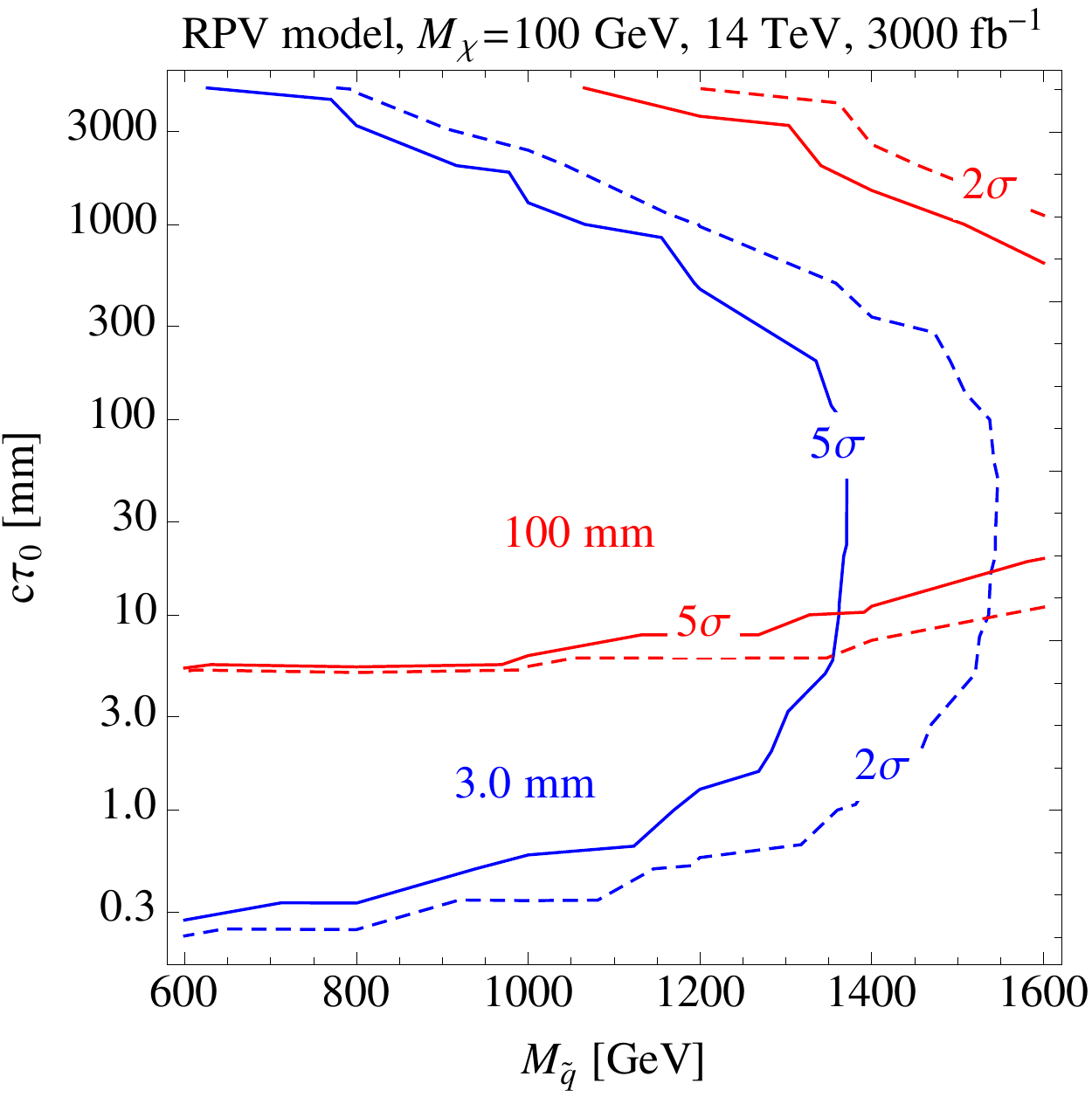}
\caption{Sensitivity of the emerging jets search for the RPV MSSM toy model, at the 14~TeV LHC. Contours are as in Fig.~\ref{fig:reachAB}. A common mass $M_{\tilde{q}}$ is assumed for first and second generation right-handed up-squarks, while all other MSSM particles are assumed to be heavy. 
}
\label{fig:RPVreach}
\end{figure}

In the following we generate events for a RPV toy model where only the right-handed up and charm squarks and the lightest neutralino are kinematically accessible. Signal events are generated using the MSSM implementation~\cite{Desai:2011su} in  \textsc{Pythia}. 
The squark masses $M_{\tilde{u}R} = M_{\tilde{c}R} \equiv M_{\tilde{q}}$ and the neutralino lifetime $c \tau_\chi$ are varied, and the neutralino mass is taken to be  $m_\chi = 100$~GeV. Since the squark masses are of order TeV, the neutralino will have a significant boost, such that its decay products will be collimated. This is a challenging regime for searches which rely on reconstructing a common displaced vertex for a dijet pair. The emerging jets search has no problem picking up this signature, and we show our reach estimate in Fig.~\ref{fig:RPVreach}. There is sensitivity across four orders of magnitude in neutralino lifetime $c\tau_0$ for squark masses as high as 1500~GeV. Compared with the dark QCD signature, the reach in $c \tau_0$ is larger. The reason for this is that there is only one displaced decay per jet, while in the dark QCD model multiple displaced decays happen, which reduce the cut efficiency on the signal. Similar to the dark QCD case, going to 3000~fb$^{-1}$ can significantly improve the reach in the 100~mm channel, while the benefits in the 3~mm search are more moderate. 

Before concluding, we would like to stress that the supersymmetric model used here was chosen purely for phenomenological reasons. From a naturalness perspective it would be more motivated to only have third generation squarks in the kinematic range. The resulting signature with prompt top-jets and displaced neutralino jets would be interesting to study in the future.

%--------------------------------------------------------------------%
\section{Conclusions}
\label{sec:conclusion}
%--------------------------------------------------------------------%

The LHC and its detectors are excellent machines for exploring the physics of the TeV scale. Yet, there are only a finite number of analyses that can be done on the data, so it is important to explore possible new physics signatures that could have been missed by current analyses. Here we have presented a new collider object, the emerging jet, which arises in many  well motivated models of physics beyond the SM, including models of dark matter which explain the coincidence between the energy density of baryons and dark matter. 

In theories with confinement in the hidden sector and a mediator much heavier than the confinement scale, there will be jet like structures produced at the LHC. If there are some long lived particles in the dark sector, a natural consequence of the separation of scales between the mediator and the hidden sector, then the dark sector jets will have large numbers of displaced vertices within them. This is a very unique experimental signature, which means most current searches will be at best very weakly sensitive to the phenomenology. 

In this work we have proposed strategies which are based on looking for signals with features that are very unlikely to be produced by QCD backgrounds. Our main method is looking for jets with very few prompt tracks. The vast majority of hard QCD jets have a large number of prompt tracks, and only very rarely do they have few or none. We have also presented an alternative strategy using $p_T$ weighting of displaced tracks. This alternative strategy is more robust to beam remnants and pile up, although it is slightly less sensitive. With the handles presented here, the LHC can be sensitive to purely hadronic signatures without missing energy that have naive signal to background ratios worse than $\mathcal{O}(10^{-3})$, and have reaches for mediator masses well above 1 TeV over several orders of magnitude in dark pion lifetimes. 

While the bulk of our analysis focuses on the general purpose detectors of the LHC, this signature also provides unique opportunities for LHCb. While LHCb does not have full coverage of the event geometry, it can be sensitive if only a few of the dark pions are within the detector geometry. Furthermore, the superior tracking of the detector needed to precisely measure $b$ hadrons can be used to precisely identify and measure dark pions and discriminate them from the zoo of QCD hadrons. Therefore, LHCb could be sensitive to a different range of dark pion masses and lifetimes than the other detectors, making it potentially the exclusive discovery machine for certain types of models.

Finally, we note that while the searches proposed here were designed with certain types of models in mind, they are potentially sensitive to a much broader classes of models with displaced phenomena including RPV SUSY and the models searched for in several current displaced analyses~\cite{CMS:2014wda,Aad:2015asa,Aaij:2014nma}. With the higher energy Run II of the LHC run about to begin, this is a great time for novel searches for new physics, and emerging jets provide an opportunity for a possible groundbreaking discovery.

\subsection*{Acknowledgments} 
We would like to thank P.~Skands for help with \textsc{Pythia}, and A.~Kagan, G.~Perez, M.~Ramsey-Musolf, G.~Salam, M.~Strassler,~M.~Tytgat, B.~Zaldivar, and J.~Zupan for useful comments and discussions. Furthermore we would like to thank the exotics conveners of the ATLAS, CMS, and LHCb collaborations for allowing us to present our work in their group meetings, and Victor Coco, Maurizio Pierini, Daniel Ventura, and Andrzej \.{Z}ura\'{n}ski for their experimental expertise. PS and DS are grateful to the Kavli Institute for Theoretical Physics and their stimulating program ``Particlegenesis'' where part of this work was completed. PS would also like to thank the Mainz Institute for Theoretical Physics (MITP) for its hospitality and support. 

\appendix

%--------------------------------------------------------------------%
\section{Collider Simulation} 
\label{sec:simulation}
%--------------------------------------------------------------------%

\subsection{Signal Events}

In the context of Hidden Valley model phenomenology~\cite{Strassler:2006im}, a dark QCD sector with $SU(N_d)$ gauge symmetry was implemented~\cite{Carloni:2010tw,Carloni:2011kk} in the event generator \textsc{Pythia}~\cite{Sjostrand:2007gs}. The model contains $n_f$ dark quarks in the fundamental representation of $SU(N_d)$ and scalar mediators of the type $X_d$ as well as the possibility to couple the dark quarks to a $Z_d$ boson. Furthermore the model implements a parton shower and fragmentation in the dark sector, with some simplifications. The string fragmentation produces only dark mesons which are either scalar (dark pion) or vector resonances (dark rho), but no dark baryons. This is a good approximation for large $N_d$ theories, but probably represents an ${\cal O}(10\%)$ error for $N_d = 3$ with a QCD-like spectrum as considered in this work. Gluon splittings into dark quark pairs are also absent.  

\begin{figure}
\centering
\includegraphics[width=1.\textwidth]{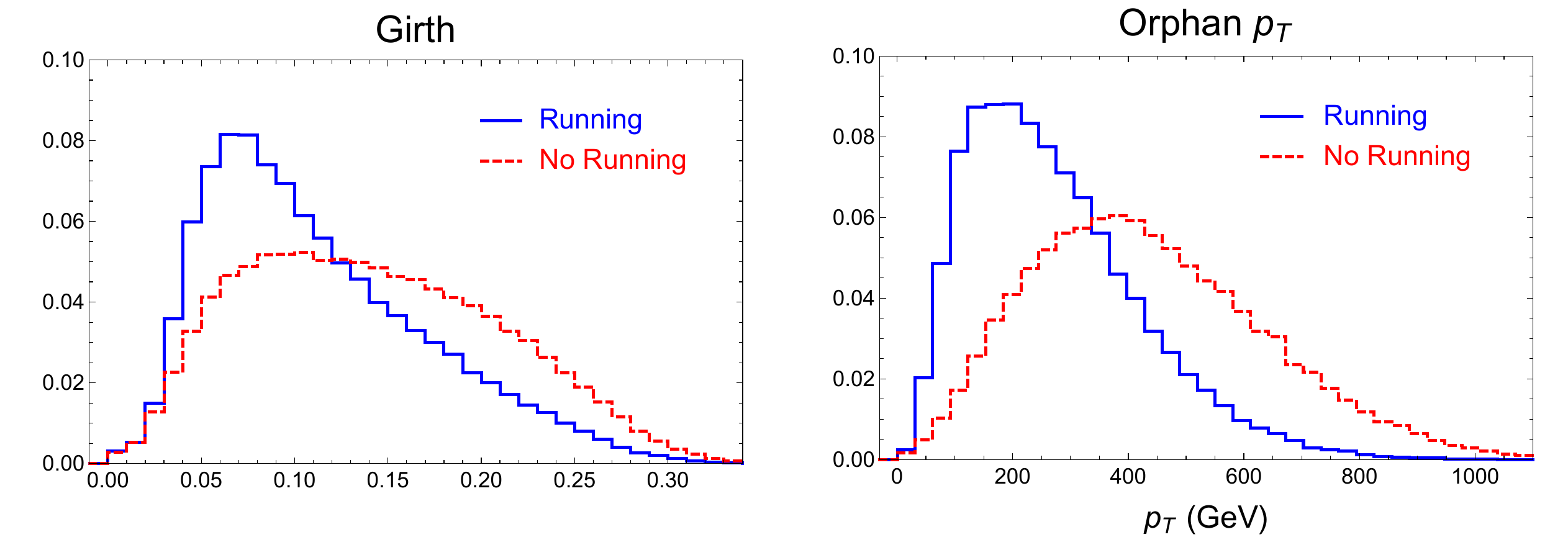}
\caption{Comparison of \textsc{Pythia} with (solid, blue) and without (dashed, red) running of the gauge coupling in the dark sector implemented (we use a coupling of 0.7 when there is no running, see text). The left plot is the girth distribution (see Eq.~(\ref{eqn:girth})), while the right plot is the orphan $p_T$: the scalar sum of the $p_T$ of visible particles which are not clustered into a jet of $p_T > 200$ GeV. This is for model B events with $Z_d$ production so all jets originate from the dark sector.  }
\label{fig:run}
\end{figure}

More importantly, the dark sector gauge coupling is not running but instead implemented as a fixed parameter, and the equivalent of the confinement scale is mimicked by introducing explicit dark quark masses.  %To test the behavior, we set the explicit quark mass equal to the confinement scale $\Lambda$ for the different benchmarks described in Table~\ref{tab:benchmarks}. We find that a gauge coupling of 0.7 roughly reproduces the total number of hadrons per event as a function of energy. 
In general, we expect that when the coupling is fixed, for large couplings events will look more spherical than in QCD-like theories, while for smaller couplings fewer particles will be produced. We can quantify this by looking at two different observables. The first is an event variable we call orphan $p_T$, which is obtained by clustering the event into jets and then summing the $p_T$ of particles which are not clustered into hard jets with $p_T > 200$ GeV. The second variable is for individual jets and is called girth~\cite{Gallicchio:2010dq}, defined as
\begin{align}
\label{eqn:girth}
{\rm girth} \, = \, \frac{1}{p_T^{\rm jet}} \sum_i p_T^i \, \Delta R_i \, ,
\end{align}
where the sum is over all constituents of the jet and $\Delta R$ is the distance in $\eta-\phi$ space of a constituent away from the jet axis. In Fig.~\ref{fig:run} we compare \textsc{Pythia} with a fixed gauge coupling of 0.7 to our modification with gauge coupling running included.\footnote{The fixed coupling of 0.7 was chosen since it most accurately reproduces the event hadron multiplicity of the case with running.} We look at events produced through a $Z_d$ so that all jets are emerging, and we see that without running, there is a lot more orphan energy and that the jets themselves tend to be broader, consistent with having events with energy spread all over the detector. 

We therefore extend the \textsc{Pythia} implementation to allow running of $\alpha_d$ from $\Lambda_{d}$ to higher scales, according to the one loop beta function with $N_d$ dark colors and $n_f$ dark flavors. As far as the phenomenology is concerned, this mainly affects the dark parton shower. It is easiest to imagine the final state parton shower\footnote{We closely follow Sec.~10 of the \textsc{Pythia} 6.4 manual~\cite{Sjostrand:2006za}.} as a series of parton branchings $a\to bc$ at scales $Q^2$.  The probability for no splitting to happen between the scales $t_0 = \log (Q_0^2/\Lambda^2)$ and $t = \log (Q^2 /\Lambda^2)$, where $\Lambda = \Lambda_d$ is the dark QCD scale here, is known as the Sudakov form factor:
\begin{align}
	{\cal P}_{a,\rm no} (t_0,t) & = \exp \left( - \int_{t^0}^t dt' \sum_{b,c} {\cal I}_{a\to bc} (t') \right) ,
	\label{eqn:sudakov}
\end{align}
where the sum runs over all possible splittings, and the integrated branching probabilities are 
\begin{align}
	{\cal I}_{a\to bc}(t) & = \int _{z_-(t)}^{z_+(t)} dz \frac{\alpha_d(t)}{2\pi} P_{a\to bc}(z), \label{eqn:branch}
\end{align}
where $z$ is the energy fraction carried by parton $b$, $E_b = z E_a$, and $P_{a\to bc} (z)$ are the splitting kernels that appear in the famous DGLAP evolution equations. In a Monte Carlo implementation of the parton shower, for a given parton with associated scale $t_0$, the task is to randomly choose the scale $t$ of the next splitting, such that it is distributed according to the splitting probability
\begin{align}
	P_{\rm split} (t) = -\frac{d}{dt}{\cal P}_{\rm no}(t_0,t) \,.
\end{align}
For a fixed $t_0$ this can be obtained using a uniformly distributed random number $x\in (0,1)$ and solving $x = {\cal P}_{\rm no}(t_0,t)$ for $t$. For negligible quark masses the boundaries of the integral in Eq.~(\ref{eqn:branch}) become independent of $t$ and we can write ${\cal I}(t) = C_{\rm emit} \alpha_d(t)/(2 \pi)$. For fixed $\alpha_d$ inverting the splitting probability is simple and one finds, using $t = \log (Q^2/\Lambda_d^2)$, 
\begin{align}
	Q^2 = Q_0^2 \;x^{\frac{2 \pi}{\alpha_d C_{\rm emit}}}\,,
\end{align}
which is independent of $\Lambda_{d}$, as expected. At one loop, the running of $\alpha_d$ is given by $\alpha(t) = (b_1 t)^{-1}$, where $b_1 = b_{1,d} = (11 C_A - 2 n_f)/(12 \pi)$ is the one-loop coefficient of the dark $SU(n_f)$ $\beta$-function. It is again possible to solve for $Q^2$ explicitly, and one obtains:
\begin{align}
\label{eqn:evolrunning}
	Q^2 = \Lambda_d^2 \; t_0^{\,x^{\frac{2 \pi b_1}{C_{\rm emit}}}} \,. 
\end{align}
We have modified the Hidden Valley shower implementation in \textsc{Pythia} such that the the running of $\alpha_d$ can be incorporated, according to Eq.~(\ref{eqn:evolrunning}). As discussed above, with a fixed coupling the parton shower does not faithfully reproduce QCD. If the coupling is small, too few dark mesons will be produced, and if the coupling is large, the events will be spherical and the partons will not be emitted in jet-like structures. 

The fragmentation process that follows the parton shower is a non-perturbative process and thus can only be modeled. Nevertheless there is some correspondence between the number of patrons that are radiated and the number of mesons that are produced, such that the average particle multiplicity as a function of the energy of the process is calculable up to an unknown normalization factor. In the next to leading high energy approximation (MLLA), it was found that 
\begin{align}
	\langle N(\hat{s}) \rangle & \propto \exp \left( \frac{1}{b_1} \sqrt{\frac{6}{\pi \alpha_s(\hat{s})}} + \left( \frac{1}{4} + \frac{5 n_f}{54 \pi b_1} \right)\log \alpha_s(\hat{s}) \right), \label{eqn:nmesontheory}
\end{align}
see e.g.~\cite{Ellis:1991qj} for a partial derivation. This behavior of the average multiplicity as a function of the energy has been verified experimentally for QCD in $e^+ e^- \to \bar{q} q$ processes. 

\begin{figure}
\center
\includegraphics[width=0.6\textwidth]{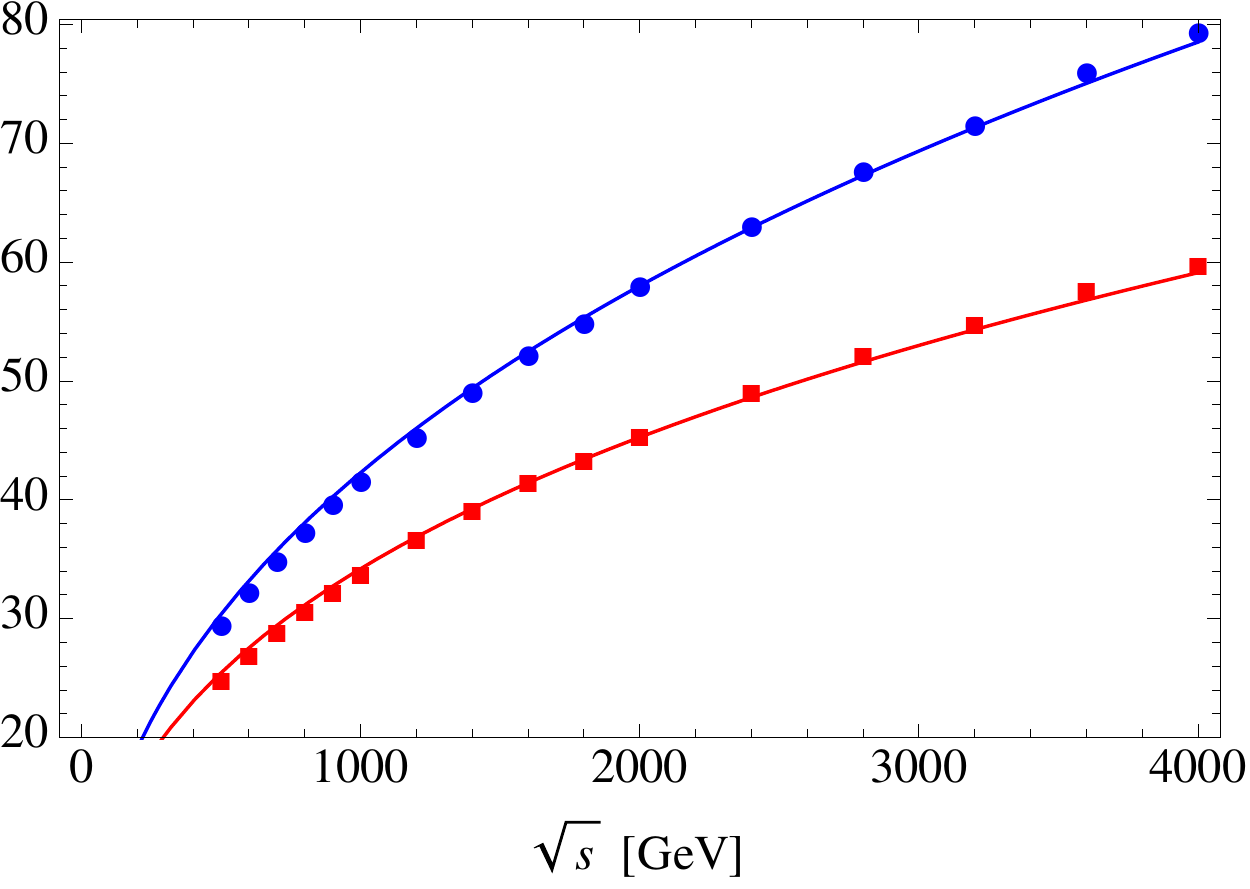}
\caption{Average dark meson multiplicity in $e^+ e^- \to Z_d^{*} \to \bar{Q}_d Q_d$ as a function of the center-of-mass energy $\sqrt{s}$. We compare the output of the modified \textsc{Pythia} implementation for $n_f = 7$ (blue circles) and $n_f = 2$ (red squares) to the theory prediction Eq.~(\ref{eqn:nmesontheory}), where we only float the normalization. The dark QCD scale and dark meson spectrum corresponds to benchmark model B. 
}
\label{fig:mesonmult}
\end{figure}

To test the modified dark QCD parton shower implementation in \textsc{Pythia}, we simulate production of dark quark pairs through a $Z_d$ boson in $e^+e^-$ collisions at center-of-mass energies between 500~GeV and 4~TeV, followed by a dark parton shower. We set the dark pions
to be stable here. The energy dependence of the average particle multiplicity is shown in Fig.~\ref{fig:mesonmult} and agrees well with the theoretical prediction Eq.~(\ref{eqn:nmesontheory}). For smaller $n_f$, the running of the coupling to smaller values is faster, so fewer partons are radiated at higher scales, resulting in a lower number of dark mesons. This is the reason for the difference in the curves for $n_f = 2$ and $n_f = 7$, and further highlights the importance of including the running coupling in the analysis.\footnote{The source for this modification of \textsc{Pythia} can be found at \url{https://github.com/pedroschwaller/EmergingJets}. }

\subsection{Modified Background Events}
\label{app:modifyPythia}

QCD backgrounds are simulated using four jet events generated in  \textsc{MadGraph5\_aMC@NLO}~\cite{Alwall:2014hca} followed by showering and hadronization in \textsc{Pythia}. The fraction of jets that are either trackless or emerging is very small, such that one might worry that the simulation is not fully accurate in this regime and might underestimate this fraction. Besides the default \textsc{Pythia} settings, we have therefore performed additional simulations with a modified tune~\cite{peter} that is designed to increase the number of jets with few mesons, to increase the probability of jets with few charged tracks, and to increase the strange components of jets, while still being marginally consistent with the low energy data that is used to tune \textsc{Pythia}. These modification increase the probability that a jet will have few prompt tracks and also increase the number of long lived states in the jet.  In the following we briefly explain how this is achieved in \textsc{Pythia}. 

Fragmentation is a nonperturbative process and in \textsc{Pythia} it is modeled using the so called Lund string fragmentation model~\cite{Andersson:1998tv} with a small number of parameters that are fit to the data. The Lund symmetric fragmentation function~\cite{Skands:2014pea}
\begin{align}
	f(z) \propto \frac{(1-z)^{a}}{z} {\rm exp} \left( \frac{-b\, m_\perp^2}{z} \right) ,
\end{align}
governs the fraction of (longitudinal) energy $z$ that is carried by a hadron which is split off from the string. Here $m_\perp^2 = m_{\rm had}^2 + p_{\perp,\rm had}^2$ is the transverse mass of the produced hadron, and $a$ and $b$ are the free parameters which are fit to the data, with default values $a=0.3$ and $b=0.8$. Larger values of $a$ reduce the probability that a large fraction of the energy is carried away by a single hadron, i.e.~the large $z$ region. Instead a larger $b$ parameter suppresses the small $z$ region. Therefore in order to increase the number of jets with only a few hard mesons, we can reduce $a$ and increase $b$. The modified tune used in the text corresponds to $a=0.26$ and $b=0.9$, and as can be seen from Fig.~\ref{fig:trackless-bg}, it leads to a slight increase in the number of background events that pass the emerging jet cuts. 

From Fig.~\ref{fig:bgComp} we see that strange mesons are the dominant background for jets emerging at distances larger than $100$~mm and contribute significantly to purely trackless jets. Therefore in order to obtain a conservative estimate for the background in that region, in our modified tune we also increase the amount of strange mesons produced in the fragmentation process by about 30\%, by changing the value of {\tt StringFlav:ProbStoUD} from 0.19 to 0.25 in \textsc{Pythia}. These parameter values are chosen to be as extreme as possible while still being marginally compatible with the soft QCD data that is used to tune \textsc{Pythia}~\cite{peter}, thereby giving a conservative upper bound on the background.

 \subsection{(Crude) Detector Simulation}
 \label{app:det-sim}

While we cannot do a full detector simulation, in this appendix we describe the way we mock up a detector to capture the key aspects necessary to capture our signal and background. Throughout this paper, we use truth level displacements and energies after hadronization.

\begin{figure}
\centering
 \includegraphics[width=.4\linewidth]{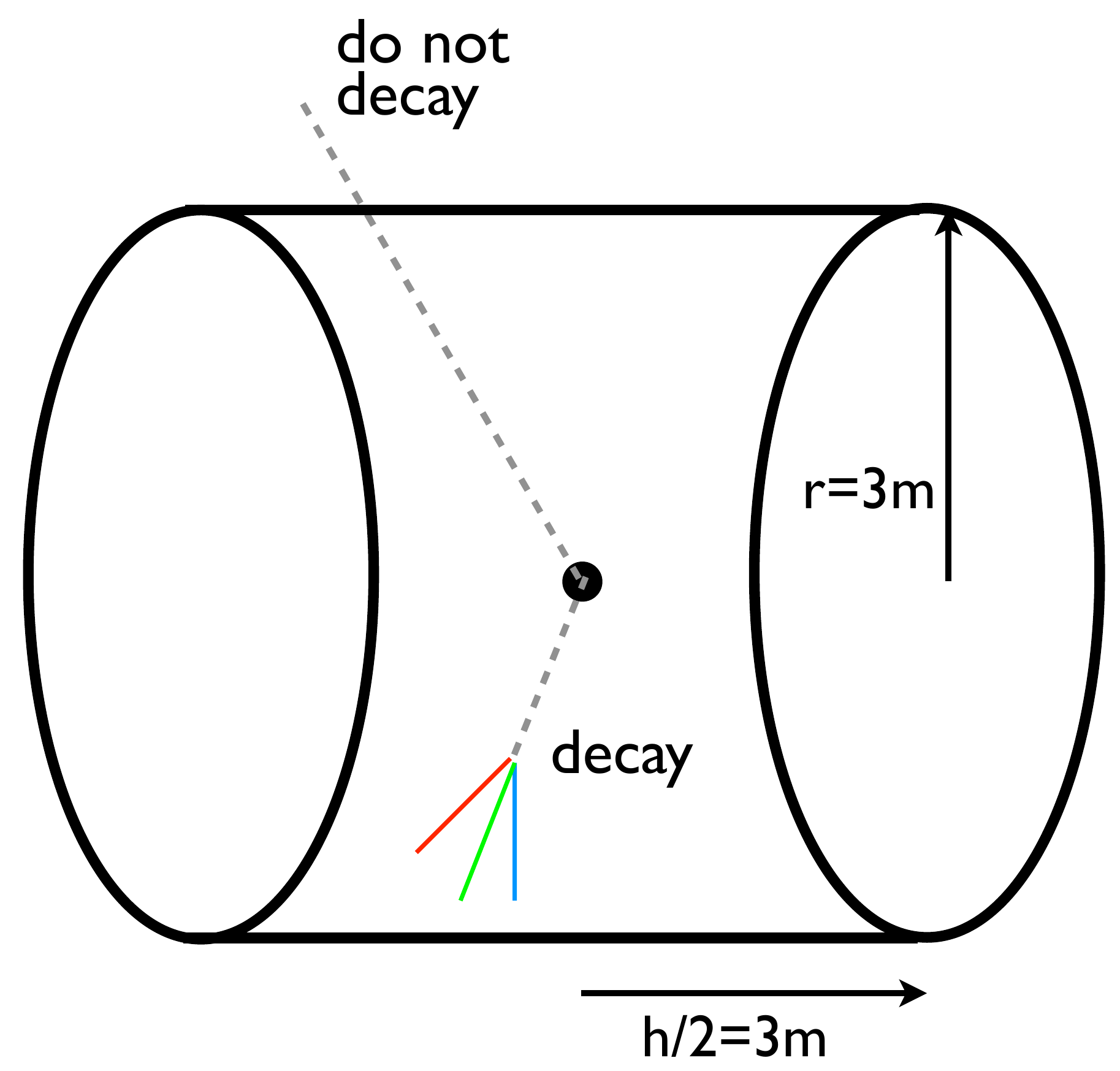}
\caption{Crude detector geometry we use: we model the calorimeter as a cylinder of radius 3 meters and height 6 meters. Particles that would decay inside the cylinder are decayed, while particles that would decay outside are left undecayed. }
\label{fig:cylinder}
\end{figure}

The first modification to truth level is to simulate the bulk geometry of the calorimeter as shown in Fig.~\ref{fig:cylinder}. We assume that dark particles which decay outside the calorimeter are not counted towards the jet energies, so we force \textsc{Pythia} not to decay particles that would have decayed outside the calorimeter. We make the simplification of a cylindrical calorimeter of radius 3 meters and height 6 meters, which is the approximate geometry of the hadronic calorimeters at both CMS and ATLAS. This cylinder effect is very important for model A with a proper lifetime of 150 mm, since in that case the majority of events have at least one undecayed dark pion. Furthermore, the pions that travel the furthest are the ones that tend to have the most energy because of relativistic boost, and this effect explains why in Fig.~\ref{fig:pt}, in model A the emerging jets tend to be softer than the standard jets in the signal events. 

The other important detector simulation comes in determining precisely how to deal with displaced particles. In QCD it is common for a charged particle to propagate through the detector and then decay to 1 (or more) charge particles. This is uncharacteristic of the signal where the long lived particles are all neutral. Therefore, we want to reject displaced particles with charged parents when possible. On the other hand, the innermost layers of tracker material are between about 50 and 100 mm at ATLAS and CMS, so if a charged particle decays without interacting with a few tracker layers, it is difficult to infer the existence of this charged particle.\footnote{Electric charge must be conserved, but a charged particle can decay to a very soft charged particle and a neutral particle, and at the LHC environment, the very soft particle is essentially invisible.} 

\begin{figure}
\centering
 \includegraphics[width=.7\linewidth]{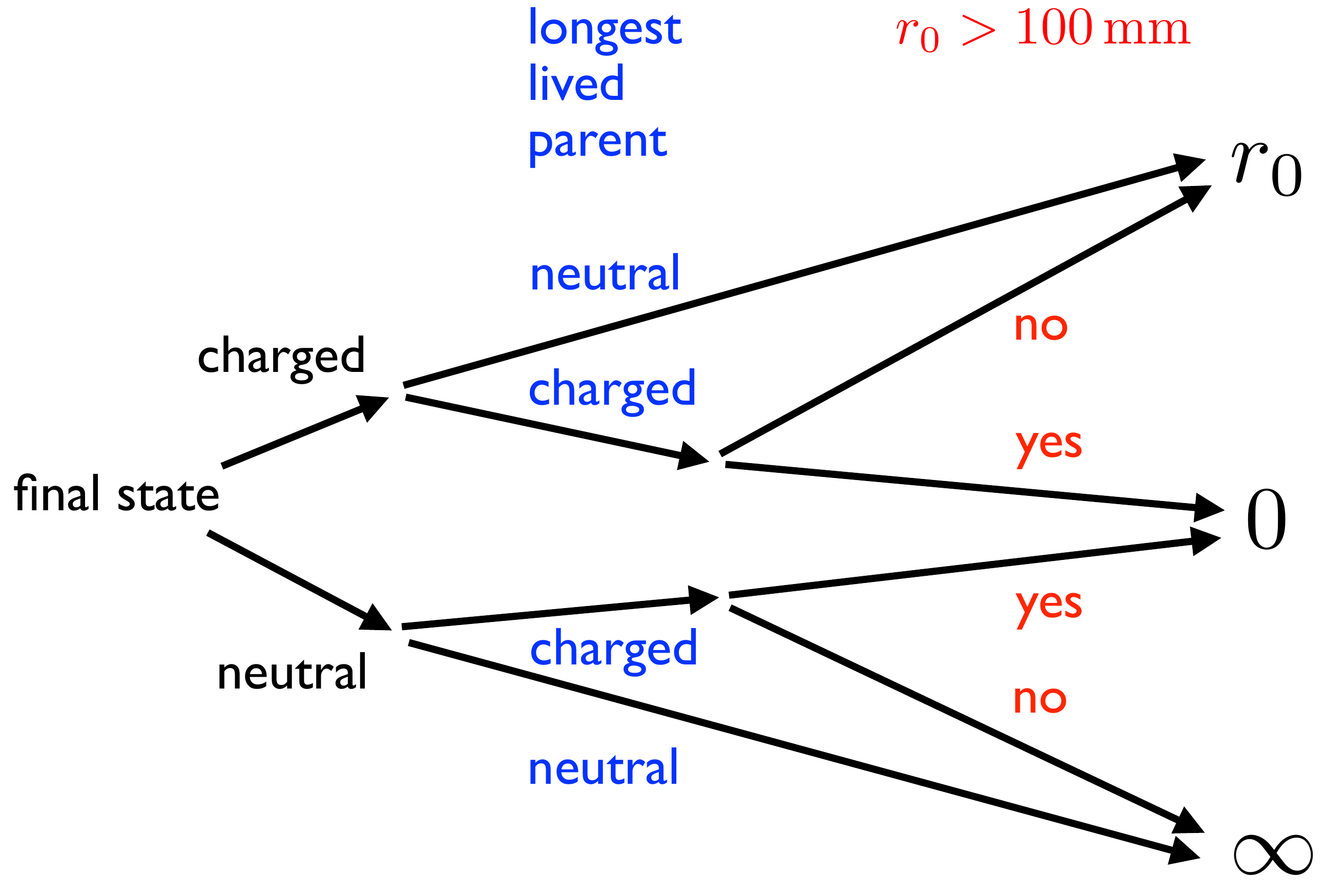}
\caption{Decision tree for determining how to assess if a particle counts as displaced or not. }
\label{fig:tree}
\end{figure}

Therefore, we implement a tracking algorithm shown schematically in Fig.~\ref{fig:tree}. We take the simplification that if a charged particle travels more than 100 mm it will be detected by the tracker. For computational simplicity, we also assume that each particle only has one long lived parent, and in the case of ambiguity we take that parent to be the one that travels the furthest in the transverse plane. Therefore, if a particle decays beyond 100 mm but it has a charged parent, it is considered prompt in the determination of the emerging property of a given jet. On the other hand, if a charged particle does not travel that far, we take it to be displaced using its truth level displacement, $r_0$ in the notation of Fig.~\ref{fig:tree}. Neutrals are ignored  (assigned a distance of infinity) for the purpose of this algorithm, unless they travel beyond 100 mm and have a charged parent. Relative to just using truth level displacement information and ignoring parentage, this reduces the signal efficiency by about 10\% and increases the background rejection by about 50\%.

%
%\begin{table}[t]
%\begin{center}
%\begin{tabular}{|c|c|c|c| }
%\hline
% Final State & Longest Lived Parent  & $r_0 >$ 100 mm & Emerging distance \\ \hline\hline
%  charged    &       charged                &          no           &       $r_0$ \\ \hline
%  charged    &       charged                &         yes           &       0    \\  \hline
%  charged    &       neutral                  &          no           &       $r_0$ \\ \hline
%  charged    &       neutral                  &         yes           &       $r_0$ \\  \hline
%  neutral      &       charged                &          no           &       $\infty$ \\ \hline
%  neutral      &       charged                &         yes           &       $0$ \\  \hline
%  neutral      &       neutral                  &          no           &       $\infty$ \\ \hline
%  neutral      &       neutral                  &         yes           &       $\infty$ \\  \hline    
%\end{tabular}
%\end{center}
%\caption{Decision table for determining emerging distance to assign to a given particle as a function of its truth level transverse displacement from the interaction point $r_0$.  }
%\label{tab:decision}
%\end{table}%

The final piece of detector realism we add concerns the background. The dominant background consists of a jet whose energy is dominated by a single photon. Because the LHC's detectors are designed to detect photons, we assume that these kinds of jets can be distinguished from the signal, and we do not count jets where at least 90\% of the energy comes from a single photon as displaced. If anything this is conservative because there are also jets with multiple photons, which can potentially be discarded by using cuts on the ratio of electromagnetic to hadronic energy, or by using information from $\gamma \to e^+e^-$ conversions. 

\begin{figure}
\centering
  \includegraphics[width=.5\linewidth]{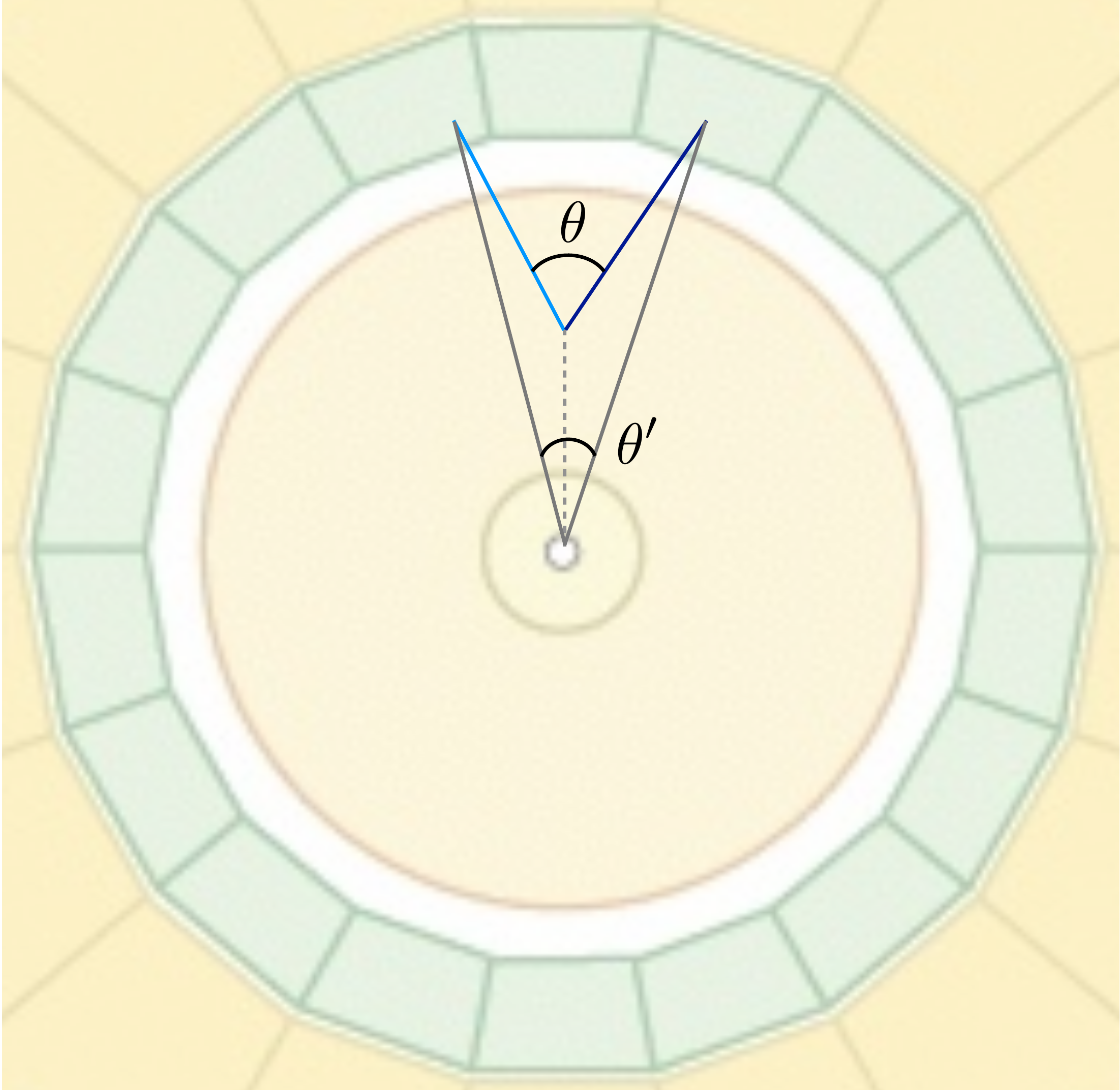}
\caption{Error in opening angle introduced by displaced vertices. Our jet algorithm uses the momentum to determine the opening angle $\theta$, which overestimates the opening angle $\theta'$ seen by the actual calorimeter.  }
\label{fig:geometry}
\end{figure}

Finally, we note one aspect of the detector simulation that we do not attempt to undertake which is in principle important, but ends up being quantitatively minor. Because we do not fully simulate the geometry of the detector, the opening angle between two final states is determined solely by their momentum vectors. On the other hand, if they originate from much closer to the calorimeter than the primary interaction vertex, we will overestimate the opening angle as shown in Fig.~\ref{fig:geometry}. This is a particularly important effect for model A with the pion lifetime being large. We can quantify this by redoing the analysis with a larger jet clustering radius, which would partially simulate capturing more of the decay products into the same jet. We find that raising the jet radius $R$ from 0.5 to 1.0 increases the energy of a typical emerging jet by 5\%, showing that this is a quantitatively unimportant effect.

%--------------------------------------------------------------------%
\section{Exploring Dark Sector Parameters} 
\label{sec:dark_sector}
%--------------------------------------------------------------------%

In the following we explore how variations of the model parameters affect the phenomenology in order to assess the model dependence of the signatures considered in this paper. The underlying theory is specified by the number of dark colors $N_d$ and the number of dark quark flavors $n_f$. We have already seen in Appendix~\ref{sec:simulation} that the number of dark mesons that are produced increases with increasing $n_f$, which happens because with larger $n_f$ the coupling runs more slowly, such that there is more radiation. However $n_f$ can not be increased arbitrarily. For $n_f \gtrsim 4 N_d$ one reaches the conformal window~\cite{Appelquist:1999hr}, where the theory runs into a fixed point in the infrared and therefore will not behave QCD like anymore. On the other end $n_f = 2$ is the minimal number of flavors that allows for proton- and neutron-like baryonic bound states. Within the range
\begin{align}
	2 \leq n_f < 4 N_d
\end{align}
it is reasonable to assume that the theory will behave similar to QCD. A change in the number of dark colors $N_d$ will have a similar effect to changing $n_f$, since both enter the $\beta$-function coefficient.\footnote{Changing $N_d$ can also affect other properties of the theory. For example for even $N_d$ the baryonic states in the theory will be bosonic. Yet the collider signature of these models is dominated by the mesons which should behave similarly.} Therefore we do not expect significant changes in the signal from variations of $N_d$ and $n_f$, as long as the parameters are chosen such that the theory is asymptotically free. The change in meson multiplicities is notable, but not large enough to invalidate our proposed search. 

Another crucial parameter is the dark confinement scale $\Lambda_d$ and the particle masses that are associated with it. We have already seen in the main part of this work that within the mass range motivated by dark matter, i.e. $\Lambda_d$ of order $1-10$~GeV, there is no strong dependence on this parameter.

\begin{figure}[tb]
\centering
\includegraphics[width=1.\textwidth]{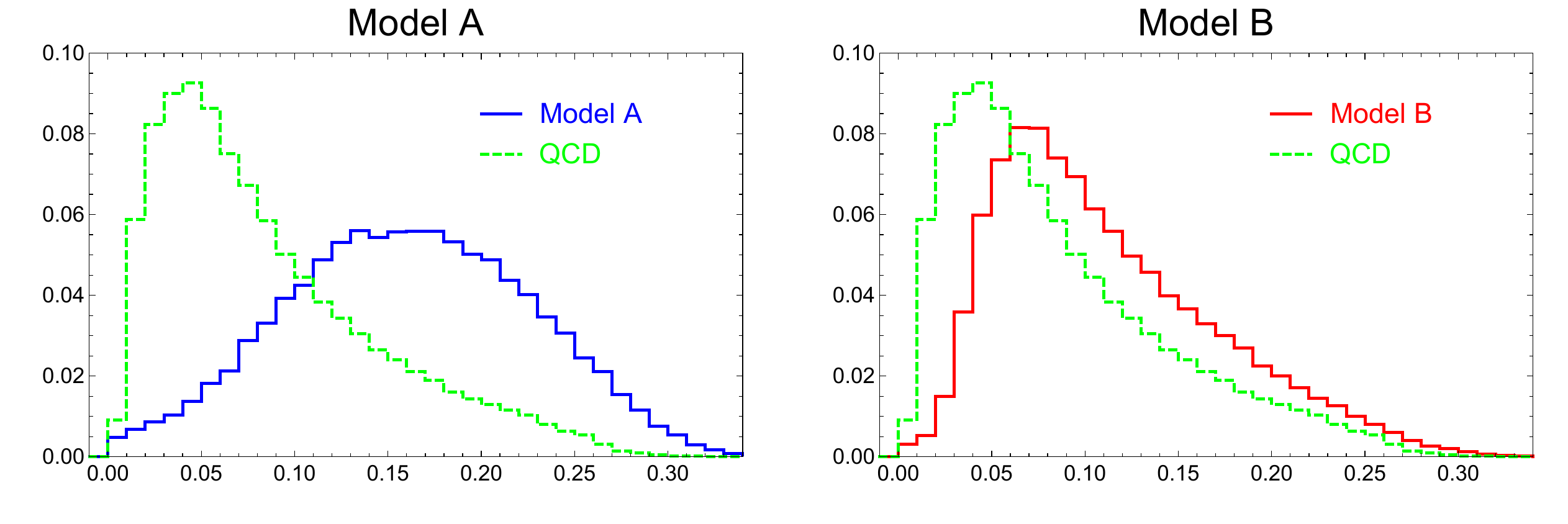}
\caption{Girth distribution for signal vs. background. The background (green, dashed) in both plots is four jet QCD events passing the kinematic cuts of Tab.~\ref{tab:cut-flow4}, while the signal are model A (left, blue, solid) and model B (right, red, solid) in the $Z_d$ model only requiring that jets have $p_T > 200$ GeV. }
\label{fig:girth}
\end{figure}

Some jet observables can, however, be sensitive to the mass scale. One such example is the girth of an individual jet defined in Eq.~(\ref{eqn:girth}). 
The distribution depends on the jet-clustering algorithm. Using the same jet parameters as in the rest of this work, we plot the girth distributions for emerging and QCD jets in Fig.~\ref{fig:girth}. For the background, we use QCD 4-jet events passing the kinematic cuts in Tab.~\ref{tab:cut-flow4}, while for the signal, we get a pure sample of emerging jets by using the $Z_d$ model and only requiring that each jet has $p_T > 200$ GeV. 

For model B, the girth distribution looks roughly like that of QCD, but for model A the difference substantial. The main reason for this is because of our detector mockup described in App.~\ref{app:det-sim}. Dark mesons which decay beyond the calorimeters are not counted towards the energy of jets. These calorimeter jets exclude the longest lived mesons, particularly in model A where the proper lifetime is 150 mm (this is a small effect in model B where $c\tau=5$ mm). The dark pions that live the longest are the ones that carry the most energy, so energetic core in of the jet will be modified in a significant way, changing the jet shapes. Without our detector simulation, the girth in model A looks much more like model B and QCD.
Therefore, in order to keep the range of validity of our search as broad as possible, we suggest not to introduce additional discriminants based on jet observables. While they could increase the sensitivity to a particular scenario, they might induce additional model dependence at the same time. 

Motivated by QCD we have considered a particle spectrum where the dark pions $\pi_d$ are parametrically lighter than other dark mesons. Instead if their masses where similar to the other dark mesons, the overall multiplicity of dark mesons would be reduced by at most a factor of about two, since the decay of heavier dark mesons to dark pions would no longer be kinematically allowed. In this scenario, however, the baryon fraction may be increased because there is no kinematic suppression for hadronizing baryons as there is in QCD. We leave a study of this scenario for future work. 

Finally one can ask how the quark flavor composition of the dark pion decays influences the signal properties. For the mass range considered here, only decays to down and strange quarks are possible. We have simulated scenarios with 100\% branching ratios into either down quarks or strange quarks, and found no significant change in the signal properties. For larger masses one should also consider decays to bottom quarks, and similarly one could also consider decays to up-type quarks instead of down-type quarks. Heavy flavors like charm and bottom quarks have a larger probability to produce muons in their decay chains, which could be useful both for triggering and signal reconstruction. However in order to keep the analysis as generic as possible, we have not considered these possibilities here.

%%%%%%%%%%%%%%%%%%%%%%%%%%%%%%%%%%%
 \end{document}